\documentclass[aps,prd,eqsecnum,showpacs,superscriptaddress,twocolumn,nofootinbib,floatfix,preprintnumbers,altaffilletter]{revtex4}
\usepackage{graphicx}
\usepackage{hyperref}
\usepackage{amssymb}
\usepackage{amsmath}
\usepackage{longtable}
\usepackage{rotating}
\usepackage{color}
\usepackage{fancyhdr}

\def\be{\begin{equation}}
\def\ee{\end{equation}}
\def\bi{\begin{itemize}}
\def\ei{\end{itemize}}
\def\ben{\begin{enumerate}}
\def\een{\end{enumerate}}

\begin{document}


\title{
Analysis of First LIGO Science Data for Stochastic Gravitational
Waves
}

%
%
%
\newcommand*{\AG}{Albert-Einstein-Institut, Max-Planck-Institut f\"ur Gravitationsphysik, D-14476 Golm, Germany}
\affiliation{\AG}
\newcommand*{\AH}{Albert-Einstein-Institut, Max-Planck-Institut f\"ur Gravitationsphysik, D-30167 Hannover, Germany}
\affiliation{\AH}
\newcommand*{\AN}{Australian National University, Canberra, 0200, Australia}
\affiliation{\AN}
\newcommand*{\CH}{California Institute of Technology, Pasadena, CA  91125, USA}
\affiliation{\CH}
\newcommand*{\DO}{California State University Dominguez Hills, Carson, CA  90747, USA}
\affiliation{\DO}
\newcommand*{\CA}{Caltech-CaRT, Pasadena, CA  91125, USA}
\affiliation{\CA}
\newcommand*{\CU}{Cardiff University, Cardiff, CF2 3YB, United Kingdom}
\affiliation{\CU}
\newcommand*{\CL}{Carleton College, Northfield, MN  55057, USA}
\affiliation{\CL}
\newcommand*{\CO}{Cornell University, Ithaca, NY  14853, USA}
\affiliation{\CO}
\newcommand*{\FN}{Fermi National Accelerator Laboratory, Batavia, IL  60510, USA}
\affiliation{\FN}
\newcommand*{\HC}{Hobart and William Smith Colleges, Geneva, NY  14456, USA}
\affiliation{\HC}
\newcommand*{\IU}{Inter-University Centre for Astronomy  and Astrophysics, Pune - 411007, India}
\affiliation{\IU}
\newcommand*{\CT}{LIGO - California Institute of Technology, Pasadena, CA  91125, USA}
\affiliation{\CT}
\newcommand*{\LM}{LIGO - Massachusetts Institute of Technology, Cambridge, MA 02139, USA}
\affiliation{\LM}
\newcommand*{\LO}{LIGO Hanford Observatory, Richland, WA  99352, USA}
\affiliation{\LO}
\newcommand*{\LV}{LIGO Livingston Observatory, Livingston, LA  70754, USA}
\affiliation{\LV}
\newcommand*{\LU}{Louisiana State University, Baton Rouge, LA  70803, USA}
\affiliation{\LU}
\newcommand*{\LE}{Louisiana Tech University, Ruston, LA  71272, USA}
\affiliation{\LE}
\newcommand*{\LL}{Loyola University, New Orleans, LA 70118, USA}
\affiliation{\LL}
\newcommand*{\MP}{Max Planck Institut f\"ur Quantenoptik, D-85748, Garching, Germany}
\affiliation{\MP}
\newcommand*{\MS}{Moscow State University, Moscow, 119992, Russia}
\affiliation{\MS}
\newcommand*{\ND}{NASA/Goddard Space Flight Center, Greenbelt, MD  20771, USA}
\affiliation{\ND}
\newcommand*{\NA}{National Astronomical Observatory of Japan, Tokyo  181-8588, Japan}
\affiliation{\NA}
\newcommand*{\NO}{Northwestern University, Evanston, IL  60208, USA}
\affiliation{\NO}
\newcommand*{\SC}{Salish Kootenai College, Pablo, MT  59855, USA}
\affiliation{\SC}
\newcommand*{\SE}{Southeastern Louisiana University, Hammond, LA  70402, USA}
\affiliation{\SE}
\newcommand*{\SA}{Stanford University, Stanford, CA  94305, USA}
\affiliation{\SA}
\newcommand*{\SR}{Syracuse University, Syracuse, NY  13244, USA}
\affiliation{\SR}
\newcommand*{\PU}{The Pennsylvania State University, University Park, PA  16802, USA}
\affiliation{\PU}
\newcommand*{\TC}{The University of Texas at Brownsville and Texas Southmost College, Brownsville, TX  78520, USA}
\affiliation{\TC}
\newcommand*{\TR}{Trinity University, San Antonio, TX  78212, USA}
\affiliation{\TR}
\newcommand*{\HU}{Universit{\"a}t Hannover, D-30167 Hannover, Germany}
\affiliation{\HU}
\newcommand*{\BB}{Universitat de les Illes Balears, E-07071 Palma de Mallorca, Spain}
\affiliation{\BB}
\newcommand*{\BR}{University of Birmingham, Birmingham, B15 2TT, United Kingdom}
\affiliation{\BR}
\newcommand*{\FA}{University of Florida, Gainsville, FL  32611, USA}
\affiliation{\FA}
\newcommand*{\GU}{University of Glasgow, Glasgow, G12 8QQ, United Kingdom}
\affiliation{\GU}
\newcommand*{\MU}{University of Michigan, Ann Arbor, MI  48109, USA}
\affiliation{\MU}
\newcommand*{\OU}{University of Oregon, Eugene, OR  97403, USA}
\affiliation{\OU}
\newcommand*{\RO}{University of Rochester, Rochester, NY  14627, USA}
\affiliation{\RO}
\newcommand*{\UW}{University of Wisconsin-Milwaukee, Milwaukee, WI  53201, USA}
\affiliation{\UW}
\newcommand*{\WU}{Washington State University, Pullman, WA 99164, USA}
\affiliation{\WU}
\author{B.~Abbott}\affiliation{\CT}
\author{R.~Abbott}\affiliation{\LV}
\author{R.~Adhikari}\affiliation{\LM}
\author{A.~Ageev}\affiliation{\MS}\affiliation{\SR}
\author{B.~Allen}\affiliation{\UW}
\author{R.~Amin}\affiliation{\FA}
\author{S.~B.~Anderson}\affiliation{\CT}
\author{W.~G.~Anderson}\affiliation{\TC}
\author{M.~Araya}\affiliation{\CT}
\author{H.~Armandula}\affiliation{\CT}
% ...
 \collaboration{The LIGO Scientific Collaboration, http://www.ligo.org}
 \noaffiliation
%
%

\date{\today}

\begin{abstract}
  We present the analysis of between 50 and 100 hrs of coincident
  interferometric strain data used to search for and establish an
  upper limit on a stochastic background of gravitational
  radiation. These data come from the first LIGO science run, during
  which all three LIGO interferometers were operated over a 2-week
  period spanning August and September of 2002. The method of
  cross-correlating the outputs of two interferometers is used for
  analysis.  We describe in detail practical signal processing issues
  that arise when working with real data, and we establish an
  observational upper limit on a $f^{-3}$ power spectrum of
  gravitational waves. Our 90\% confidence limit is
  $\Omega_0\,h_{100}^2\le 23$ in the frequency band 40 to 314~Hz, where
  $h_{100}$ is the Hubble constant in units of 100~km/sec/Mpc and
  $\Omega_0$ is the gravitational wave energy density per logarithmic
  frequency interval in units of the closure density. This limit is
  approximately $10^4$ times better than the previous, broadband
  direct limit using interferometric detectors, and nearly 3 times
  better than the best narrow-band bar detector limit.  As LIGO and
  other worldwide detectors improve in sensitivity and attain their
  design goals, the analysis procedures described here should lead to
  stochastic background sensitivity levels of astrophysical interest.

\end{abstract}

\pacs{04.80.Nn, 04.30.Db, 95.55.Ym, 07.05.Kf, 02.50.Ey, 02.50.Fz,
98.70.Vc} 

\begin{center}
\end{center}

\maketitle

\section{INTRODUCTION}
\label{sec:intro}

In the last few years a number of new gravitational wave
detectors, using long-baseline laser interferometry, have begun
operation. These include the Laser Interferometer Gravitational
Wave Observatory (LIGO) detectors located in Hanford, WA and
Livingston, LA \cite{ligoproject}; the GEO-600 detector near
Hannover, Germany \cite{GEO-600}; the VIRGO detector near Pisa,
Italy \cite{virgo}; and the Japanese TAMA-300 detector in Tokyo
\cite{tama}. While all of these instruments are still being
commissioned to perform at their designed sensitivity levels, many
have begun making dedicated data collecting runs and performing
gravitational wave search analyses on these data.

In particular, from 23 August 2002 to 9 September 2002, the LIGO
Hanford and LIGO Livingston Observatories (LHO and LLO) collected
coincident science data; this first scientific data run is
referred to as S1. The LHO site contains two identically oriented
interferometers: one having 4 km long measurement arms (referred
to as H1), and one having 2 km long arms (H2); the LLO site
contains a single, 4 km long interferometer (L1). GEO-600 also
took data in coincidence with the LIGO detectors during that time.
Members of the LIGO Scientific Collaboration have been analyzing
these data to search for gravitational wave signals. These initial
analyses are aimed at developing the search techniques and
machinery, and at using these fundamentally new instruments to
tighten upper limits on gravitational wave sources. Here we report
on the methods and results of an analysis performed on the LIGO
data to set an upper limit on a stochastic background of
gravitational waves.%
\footnote{Given the GEO S1 sensitivity level and
large geographical separation of the GEO-600 and LIGO detectors,
it was not profitable to include GEO-600 data in this analysis.}
This represents the first such analysis performed on data from
these new long-baseline detectors. The outline of the paper is as
follows:

Section II gives a description of the LIGO instruments and a
summary of their operational characteristics during the S1 data
run. In Sec.~III, we give a brief description of the properties of
a stochastic background of gravitational radiation, and Sec.~IV
reviews the basic analysis method of cross-correlating the outputs
of two gravitational wave detectors.

In Sec.~V, we discuss in detail the analysis performed on the LIGO
data set. In applying the basic cross-correlation technique to
real detector data, we have addressed some practical issues and
performed some additional analyses that have not been dealt with
previously in the literature: (i) avoidance of spectral leakage in
the short-time Fourier transforms of the data; (ii) a procedure
for identifying and removing narrow-band (discrete frequency)
correlations between detectors; (iii) chi-squared and time shift
analyses, designed to explore the frequency domain character of
the cross correlations.

In Sec.~VI, the error estimation is presented, and in Sec.~VII, we
show how the procedure has been tested by analyzing data that
contain an artificially injected, simulated stochastic background
signal. Section VIII discusses in more detail the instrumental
correlation that is observed between the two Hanford
interferometers (H1 and H2), and Sec.~IX concludes the paper with
a brief summary and topics for future work.

Appendix~\ref{sec:listofsymbols} gives a list of symbols used in
the paper, along with their descriptions and equation numbers or
sections in which they were defined.

\section{THE LIGO DETECTORS}
\label{sec:instrument}

An interferometric gravitational-wave detector attempts to
measure oscillations in the space-time metric, utilizing the
apparent change in light travel time induced by a gravitational
wave. At the core of each LIGO detector is an orthogonal arm
Michelson laser interferometer, as its geometry is well-matched
to the space-time distortion. During any half-cycle of the
oscillation, the quadrupolar gravitational-wave field increases
the light travel time in one arm and decreases it in the other
arm. Since the gravitational wave produces the equivalent of a
strain in space, the travel time change is proportional to the
arm length, hence the long arms. Each arm contains two test
masses, a partially transmitting mirror near the beam splitter
and a near-perfect reflector at the end of the arm. Each such
pair is oriented to form a resonant Fabry-Perot cavity, which
further increases the strain induced phase shifts by a factor
proportional to the cavity finesse. An additional partially
transmitting mirror is placed in the input path to form the
power-recycling cavity, which increases the power incident on the
beam splitter, thereby decreasing the shot-noise contribution to
the signal-to-noise ratio of the gravitational-wave signal.

Each interferometer is illuminated with a medium power Nd:YAG
laser, operating at 1.06 microns \cite{ligolaser}. Before the
light is launched into the interferometer, its frequency,
amplitude and direction are all stabilized, using a combination
of active and passive stabilization techniques. To isolate the
test masses and other optical elements from ground and acoustic
vibrations, the detectors implement a combination of passive and
active seismic isolation systems \cite{seismicstack,microff}, from
which the mirrors are suspended as pendulums. This forms a coupled
oscillator system with high isolation for frequencies above
$~40$~Hz. The test masses, major optical components, vibration
isolation systems, and main optical paths are all enclosed in a
high vacuum system.

Various feedback control systems are used to keep the multiple
optical cavities tightly on resonance \cite{lengthpaper} and
well-aligned \cite{alignmentpaper}. The gravitational wave strain
signal is obtained from the error signal of the feedback loop used
to control the differential motion of the interferometer arms. To
calibrate the error signal, the effect of the feedback loop gain
is measured and divided out, and the response $\widetilde R(f)$
to a differential arm strain is measured and factored in. For the
latter, the absolute scale is established using the laser
wavelength, and measuring the mirror drive signal required to
move through a given number of interference fringes. During
interferometer operation, the calibration was tracked by injecting
fixed-amplitude sinusoidal signals into the differential arm
control loop, and monitoring the amplitude of these signals at
the measurement (error) point \cite{gaby-cal}.

\begin{figure}[htbp!]
  \includegraphics[width=3.5in,angle=0]{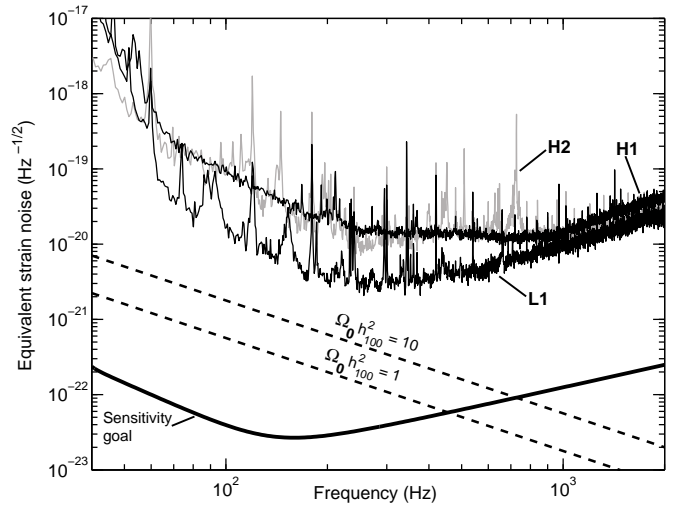}
  \caption{Reference sensitivity curves for the three LIGO interferometers
  during the S1 data run, in terms of equivalent strain noise
  density. The H1 and H2 spectra are from 9 September 2002, and the L1
  spectrum is from 7 September 2002.
  Also shown are strain spectra corresponding to two levels of a stochastic
  background of gravitational radiation defined by Eq.~\ref{e:sbstrain}.
  These can be compared to the expected 90\% confidence level upper
  limits, assuming Gaussian uncorrelated detector noise at the levels
  shown here, for the interferometer pairs: 
  H2-L1 ($\Omega_0\,h_{100}^2=10$),
  with 100 hours of correlated integration time; 
  H1-H2 ($\Omega_0\,h_{100}^2=0.83$), 
  with 150 hours of integration time; and
  H1-L1 ($\Omega_0\,h_{100}^2=11$), 
  with 100 hours of integration time. 
  Also shown is the strain noise goal for the two 4-km arm 
  interferometers (H1 and L1).
  }
  \label{fig:sensitivityPlots}
\end{figure}

Figure \ref{fig:sensitivityPlots} shows reference amplitude
spectra of equivalent strain noise, for the three LIGO
interferometers during the S1 run. The eventual strain noise goal
is also indicated for comparison. The differences among the
three spectra reflect differences in the operating parameters and
hardware implementations of the three instruments; they are in
various stages of reaching the final design configuration. For
example, all interferometers operated during S1 at a substantially
lower effective laser power level than the eventual level of 6~W
at the interferometer input; the resulting reduction in
signal-to-noise ratio is even greater than the square-root of the
power reduction, because the detection scheme is designed to be
efficient only near the design power level. Thus the shot-noise
region of the spectrum (above $~200$~Hz) is much higher than the
design goal. Other major differences between the S1 state and the
final configuration were: partially implemented laser frequency
and amplitude stabilization systems; and partially implemented
alignment control systems.

Two other important characteristics of the instruments'
performance are the stationarity of the noise, and the duty cycle
of operation. The noise was significantly non-stationary, due to
the partial stabilization and controls mentioned above. In the
frequency band of most importance to this analysis, approximately
$60-300$~Hz, a factor of $~2$ variation in the noise amplitude
over several hours
was typical for the instruments; this is addressed quantitatively
in Sec.~\ref{sec:errors} and Fig.~\ref{fig:s1weights}. As 
our analysis relies on cross-correlating the outputs of two
detectors, the relevant duty cycle measures are those for
double-coincident operation. For the S1 run, the total times of
coincident science data for the three pairs are: H1-H2, 188~hrs
(46\% duty cycle over the S1 duration); H1-L1, 116~hrs (28\%);
H2-L1, 131~hrs (32\%). A more detailed description of the LIGO
interferometers and their performance during the S1 run can be
found in reference \cite{s1ligo}.

\section{STOCHASTIC GRAVITATIONAL WAVE BACKGROUNDS}
\label{sec:sgwbs}

\subsection{Spectrum}
\label{sec:spectrum}

A stochastic background of gravitational radiation is analogous to
the cosmic microwave background radiation, though its spectrum is
unlikely to be thermal. Sources of a stochastic background could
be cosmological or astrophysical in origin. Examples of the former
are zero-point fluctuations of the space-time metric amplified
during inflation, and first-order phase transitions and decaying
cosmic string networks in the early universe. An example of an
astrophysical source is the random superposition of many weak
signals from binary-star systems. See references \cite{leshouches}
and \cite{maggiore} for a review of sources.

The spectrum of a stochastic background is usually described by
the dimensionless quantity $\Omega_{\rm gw}(f)$ which is the
gravitational-wave energy density per unit logarithmic frequency,
divided by the critical energy density
$\rho_{\rm c}$ to close the universe:
\begin{equation}
\Omega_{\rm gw}(f)\equiv \frac{f}{\rho_{\rm c}}\,
\frac{d\rho_{\rm gw}}{df}\ .
\label{e:Omega_gw}
\end{equation}
The critical density $\rho_{\rm c} \equiv 3 c^2 H_0^2/ 8 \pi G$
depends on the present day Hubble expansion rate $H_0$.  For
convenience we define a dimensionless factor
\begin{equation}
h_{100} \equiv H_0/H_{100}\,,
\label{e:h_100}
\end{equation}
where
\begin{equation}
H_{100} \equiv 100\ {\rm  km\over sec\cdot Mpc} \approx
3.24 \times 10^{-18}\ {\rm 1 \over sec}\,,
\label{e:H_100}
\end{equation}
to account for the different values of $H_0$ that are quoted in 
the literature.%
\footnote{$H_0 = 73\pm2\pm7$~km/sec/Mpc as shown in Ref.~\cite{H0_1}
and from independent SNIa observations from observatories on the
ground \cite{H0_2}.
The Wilkinson Microwave Anisotropy Probe 1st year (WMAP1)
observation has $H_0 = 71^{+4}_{-3}$~km/sec/Mpc \cite{WMAP}.}
Note that $\Omega_{\rm gw}(f)\,h_{100}^2$ is independent of the
actual Hubble expansion rate, so we work with this quantity rather 
than $\Omega_{\rm gw}(f)$ alone.

Our specific interest is the measurable one-sided power spectrum
of the gravitational wave strain $S_{\rm gw}(f)$, which is 
normalized according to:
\begin{equation}
\lim_{T\rightarrow\infty}{1\over T}\int_{-T/2}^{T/2} dt\,|h(t)|^2\
= \int_0^\infty\,df\,S_{\rm gw}(f)\ ,
\end{equation}
where $h(t)$ is the strain in a single detector due to the
gravitational wave signal; $h(t)$ can be expressed in terms of the
perturbations $h_{ab}$ of the spacetime metric and the detector
geometry via:
\begin{equation}
h(t)\equiv h_{ab}(t,\vec x_0)\,
\frac{1}{2}(\hat X^a\hat X^b-\hat Y^a\hat Y^b)\ .
\label{e:h(t)}
\end{equation}
Here $\vec x_0$ specifies the coordinates of the interferometer
vertex, and $\hat X^a$, $\hat Y^a$ are unit vectors pointing in the
direction of the detector arms. Since the energy density in
gravitational waves involves a product of time derivatives of the
metric perturbations (c.f.\ p.\ 955 of Ref.~\cite{MTW}), one can show
(see, e.g., Secs.~II.A and III.A in Ref.~\cite{allenromano} for more
details) that $S_{\rm gw}(f)$ is related to $\Omega_{\rm gw}(f)$ via:
\begin{equation}
S_{\rm gw}(f)= \frac{3 H_0^2}{10\pi^2}\,f^{-3}\Omega_{\rm gw}(f)\,. 
\label{e:gwpsd}
\end{equation}
Thus, for a stochastic gravitational wave background with $\Omega_{\rm
gw}(f)\equiv\Omega_0={\rm const}$ (as is predicted at LIGO frequencies
e.g., by inflationary models in the infinitely slow-roll limit, or by
cosmic string models \cite{caldwell}) the power in gravitational waves
falls off as $1/f^3$, with a strain amplitude scale of:
\begin{eqnarray}
\lefteqn{  S_{\rm gw}^{1/2}(f) = } \nonumber \\
 & 5.6 \times 10^{-22}\ h_{100} \sqrt{\Omega_0} \left(\frac{{\rm
100\, Hz}}{f}\right)^{3/2} {\rm Hz}^{-1/2}. \label{e:sbstrain}
\end{eqnarray}

The spectrum $\Omega_{\rm gw}(f)$ completely specifies the
statistical properties of a stochastic background of gravitational
radiation provided we make several additional assumptions. Here,
we assume that the stochastic background is isotropic,
unpolarized, stationary, and Gaussian. Anisotropic or non-Gaussian
backgrounds (e.g., due to an incoherent superposition of
gravitational waves from a large number of unresolved white dwarf
binary star systems in our own galaxy, or a ``pop-corn" stochastic
signal produced by gravitational waves from supernova
core-collapse events \cite{coward,ferrari}) may require different
data analysis techniques from those presented here. (See, e.g.,
\cite{allenottewill,drasco} for a discussions of these different
techniques.)

\subsection{Prior observational constraints}
\label{sec:observational_constraints}

While predictions for $\Omega_{\rm gw}(f)$ from cosmological
models can vary over many orders of magnitude, there are several
observational results that place interesting upper limits on
$\Omega_{\rm gw}(f)$ in various frequency bands.
Table~\ref{tab:upperlimits} summarizes these observational
constraints and upper limits on the energy density of a stochastic
gravitational wave background. The high degree of isotropy
observed in the cosmic microwave background radiation (CMBR)
places a strong constraint on $\Omega_{\rm gw}(f)$ at very low
frequencies \cite{turner}. Since $H_{100}\approx 3.24\times
10^{-18}$~Hz, this limit applies only over several decades
of frequency $10^{-18}-10^{-16}$~Hz which are far below the
bands accessible to investigation by either Earth-based
($10-10^4$~Hz) or space-based ($10^{-4}-10^{-1}$~Hz)
detectors.

Another observational constraint comes from nearly two decades of
monitoring the time-of-arrival jitter of radio pulses from a
number of millisecond pulsars \cite{mchugh}. These pulsars are
remarkably stable clocks, and the regularity of their pulses
places tight constraints on $\Omega_{\rm gw}(f)$ at frequencies
on the order of the inverse of the observation time of the
pulsars, $1/T \sim 10^{-8}$ Hz. Like the constraint derived from
the isotropy of the CMBR, the millisecond pulsar timing
constraint applies to an observational frequency band much lower
than that probed by Earth-based and space-based detectors.

The only constraint on $\Omega_{\rm gw}(f)$ within the
frequency band of Earth-based detectors comes from the observed
abundances of the light elements in the universe, coupled with
the standard model of big-bang nucleosynthesis \cite{kolbturner}.
For a narrow range of key cosmological parameters, this model is
in remarkable agreement with the elemental observations. One of
the constrained parameters is the expansion rate of the universe
at the time of nucleosynthesis, thus constraining the energy
density of the universe at that time. This in turn constrains the
energy density in a cosmological background of gravitational
radiation (non-cosmological sources of a stochastic background, e.g., 
from a superposition of supernovae signals, are not of course 
constrained by these observations).
The observational constraint is on the logarithmic integral over 
frequency of $\Omega_{\rm gw}(f)$. 

All the above constraints were indirectly inferred via
electromagnetic observations. There are a few, much weaker
constraints on $\Omega_{\rm gw}(f)$ that have been set by
observations with detectors directly sensitive to gravitational
waves. The earliest such measurement was made with room-temperature 
bar detectors, using a split bar technique for wide bandwidth 
performance \cite{glasgowbar}. Later measurements
include an upper limit from a correlation between the Garching
and Glasgow prototype interferometers \cite{garchingglasgow},
several upper limits from observations with a single cryogenic
resonant bar detector \cite{astone1996,astone1999a}, and most
recently an upper limit from observations of two-detector
correlations between the Explorer and Nautilus cryogenic resonant
bar detectors \cite{astone1999b,astoneGWDAW}. Note that the
cryogenic resonant bar observations are constrained to a very narrow
bandwidth ($\Delta \rm{f}\sim 1 \rm{Hz}$) around the resonant
frequency of the bar.

\begin{table*}
\begin{ruledtabular}
\begin{tabular}{cccc}

Observational & Observed & Frequency & Comments\\
Technique & Limit & Domain &\\
\hline

Cosmic Microwave&&& \\
\rule[-4mm]{0mm}{3mm}Background & $\Omega_{\rm gw}(f)\  h_{100}^2
\le 10^{-13} \left(\frac{10^{-16}\ {\rm Hz}}{f}\right)^2 $ & $3
\times 10^{-18} {\rm Hz}< f <
10^{-16}$~Hz & \cite{turner}\\
Radio Pulsar&&&\\
\rule[-4mm]{0mm}{3mm}Timing&$\Omega_{\rm gw}(f)\ h_{100}^2 \le 
9.3\times 10^{-8}$&
$4\times 10^{-9}~{\rm Hz}< f < 4 \times 10^{-8}~{\rm Hz}$
&95\% CL bound, \cite{mchugh}\\
Big-Bang&&&\\
\rule[-3mm]{0mm}{2mm}Nucleosynthesis&$\int_{f>10^{-8}\ {\rm Hz}}
d\ln f\ \Omega_{\rm gw}(f)\ h_{100}^2
\le 10^{-5} $&$ f > 10^{-8}\rm{Hz}$&95\% CL bound,  \cite{kolbturner}\\
\hline

\rule[-4mm]{0mm}{5mm}Interferometers&$\Omega_{\rm gw}(f)\ h_{100}^2 \le 3 \times 10^{5}$&$100~\rm{Hz} \alt f \alt 1000~\rm{Hz}$
&Garching-Glasgow \cite{garchingglasgow}\\
Room Temperature&&&\\
\rule[-4mm]{0mm}{3mm}Resonant Bar (correlation)&$\Omega_{\rm
gw}(f_0)\  h_{100}^2 \le 3000$
&$f_0 =985\pm 80~\rm{Hz}$&Glasgow \cite{glasgowbar}\\
Cryogenic Resonant&$\Omega_{\rm gw}(f_0)\  h_{100}^2 \le 300$&$f_0 = 907~\rm{Hz}$ & Explorer \cite{astone1996}\\
\rule[-4mm]{0mm}{3mm}Bar (single)&$\Omega_{\rm gw}(f_0)\
h_{100}^2 \le 5000$
& $f_0 = 1875~\rm{Hz}$ & ALTAIR \cite{astone1999a}\\
Cryogenic Resonant&&&\\
Bar (correlation)&$\Omega_{\rm gw}(f_0)\  h_{100}^2 \le 60$ &
$f_0 = 907~\rm{Hz}$
&Explorer+Nautilus \cite{astone1999b,astoneGWDAW}\\
\end{tabular}
\end{ruledtabular}
\caption{Summary of upper limits on $\Omega_0\,h_{100}^2$ over a
large range of frequency bands. The upper portion of the table
lists indirect limits derived from astrophysical observations. The
lower portion of the table lists limits obtained from prior direct
gravitational wave measurement.}
\label{tab:upperlimits}
\end{table*}

\section{DETECTION VIA CROSS-CORRELATION}
\label{sec:ccstatistic}

We can express the equivalent strain output $s_i(t)$ of each of our
detectors as:
\begin{equation}
s_i(t) \equiv h_i(t) + n_i(t)\,,
\label{e:s_i(t)}
\end{equation}
where $h_i(t)$ is the strain signal in the {\em i\/}-th detector due
to a gravitational wave background, and $n_i(t)$ is the detector's
equivalent strain noise. If we had only one detector, all we could do
would be to put an upper limit on a stochastic background at the
detector's strain noise level; e.g., using L1 we could put a limit of
$\Omega_0\,h_{100}^2 \sim 10^3$ in the band $100-200$~Hz. To do much
better, we {\em cross-correlate\,} the outputs of two detectors,
taking advantage of the fact that the sources of noise $n_i$ in each
detector will, in general, be independent
\cite{mich,chris,flan,leshouches,allenromano,maggiore}. We thus
compute the general cross-correlation:%
\footnote{The equations in this section are a summary of Sec.~III from
Ref.~\cite{allenromano}.  
Readers interested in more details and/or derivations of the key
equations should refer to \cite{allenromano} and references contained 
therein.}
\begin{equation}
Y \equiv \int_{-T/2}^{T/2} dt_1\int_{-T/2}^{T/2} dt_2\,
s_1(t_1)\, Q(t_1-t_2)\, s_2(t_2)\,,
\label{e:Y_continuous_time}
\end{equation}
where $Q(t_1-t_2)$ is a (real) filter function, which we will choose
to maximize the signal-to-noise ratio of $Y$.  Since
the optimal choice of $Q(t_1-t_2)$ falls off rapidly for time delays
$|t_1-t_2|$ large compared to the light travel time $d/c$ between the
two detectors,%
\footnote{The light travel time $d/c$ between the Hanford and
Livingston detectors is approximately 10~msec.} 
and since a typical
observation time $T$ will be much, much greater than $d/c$, we can
change the limits on one of the integrations from $(-T/2, T/2)$ to
$(-\infty,\infty)$, and subsequently obtain \cite{allenromano}:
\begin{equation}
Y\approx\int_{-\infty}^{\infty}df\ \int_{-\infty}^{\infty} df'\
\delta_T(f-f')\,\tilde s_1^*(f)\,\widetilde Q(f')\,\tilde s_2(f')\,, 
\label{e:Y_continuous_freq}
\end{equation}
where
\begin{equation}
\delta_T(f)\equiv\int_{-T/2}^{T/2} dt\, e^{-i2\pi ft}
=\frac{\sin(\pi f T)}{\pi f}
\label{e:delta_T}
\end{equation}
is a finite-time approximation to the Dirac delta function, and 
$\widetilde s_i(f)$, $\widetilde Q(f)$ denote the Fourier transforms
of $s_i(t)$, $Q(t)$---i.e., 
$\widetilde a(f)\equiv\int_{-\infty}^\infty dt\, e^{-i2\pi ft}\,a(t)$.

\begin{figure}[htbp!]
  \includegraphics[width=3.5in,angle=0]{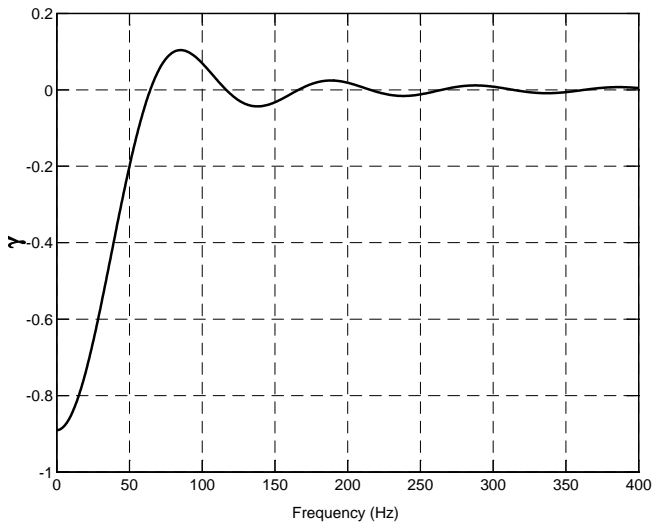}
  \caption{Overlap reduction function between the LIGO Livingston and the
    LIGO Hanford sites. The value of $|\gamma|$ is a little less than
    unity at 0~Hz because the interferometer arms are not exactly
    co-planar and co-aligned between the two sites.}
  \label{fig:LLOLHOgamma}
\end{figure}

To find the optimal $\widetilde Q(f)$, we assume that the intrinsic
detector noise is: (i) stationary over a measurement time $T$; (ii)
Gaussian; (iii) uncorrelated between different detectors; (iv)
uncorrelated with the stochastic gravitational wave signal; and (v) much
greater in power at any frequency than the stochastic gravitational
wave background. Then the expected value of the cross-correlation $Y$
depends only on the stochastic signal:
\begin{equation}
\mu_Y \equiv \langle Y \rangle =
\frac{T}{2}\int_{-\infty}^\infty df\, \gamma(|f|)
S_{\rm gw}(|f|)\, \widetilde Q(f)\,, 
\label{e:mu_continuous}
\end{equation}
while the variance of $Y$ is dominated by the noise in the
individual detectors:
\begin{equation}
\sigma_Y^2 \equiv \langle (Y - \langle Y \rangle)^2 \rangle 
\approx \frac{T}{4}\int_{-\infty}^\infty df\, 
P_1(|f|)\,|\widetilde Q(f)|^2\,P_2(|f|)\,. 
\label{e:sigma2_continuous}
\end{equation}
Here $P_1(f)$ and $P_2(f)$ are the one-sided strain noise power
spectra of the two detectors. The integrand of Eq.~\ref{e:mu_continuous} 
contains a (real) function $\gamma(f)$, called the {\em overlap reduction
function\,} \cite{flan}, which characterizes the reduction in
sensitivity to a stochastic background arising from the separation
time delay and relative orientation of the two detectors. It is a
function of only the relative detector geometry (for coincident and
co-aligned detectors, like H1 and H2, $\gamma(f)=1$ for all
frequencies). A plot of the overlap reduction function for
correlations between LIGO Livingston and LIGO Hanford is shown in
Fig.~\ref{fig:LLOLHOgamma}.

From Eqs.~(\ref{e:mu_continuous}) and (\ref{e:sigma2_continuous}), it
is relatively straightforward to show \cite{leshouches} that the
expected signal-to-noise ratio ($\mu_Y/\sigma_Y$) of $Y$ is maximized 
when
\begin{equation}
\widetilde Q(f) \propto
\frac{\gamma(|f|) S_{\rm gw}(|f|)}{P_1(|f|)P_2(|f|)}\,.
\end{equation}
For the S1 analysis, we specialize to the case 
$\rm\Omega_{\rm gw}(f)\equiv\Omega_0={\rm const}$. Then,
\begin{equation}
\widetilde Q(f) = {\cal N} \frac{\gamma(|f|)}{|f|^3 P_1(|f|)P_2(|f|)}\,,
\label{e:optimal_continuous}
\end{equation}
where ${\cal N}$ is a (real) overall normalization constant. In
practice we choose ${\cal N}$ so that the expected
cross-correlation is $\mu_Y=\Omega_0\,h_{100}^2\, T$.
For such a choice,
\begin{eqnarray}
{\cal N} &=& 
\frac{20\pi^2}{3 H_{100}^2}\
\left[\int_{-\infty}^\infty df\, 
\frac{\gamma^2(|f|)}{f^6 P_1(|f|)P_2(|f|)}
\right]^{-1}\,,
\label{e:normalization_continuous}\\
\sigma^2_Y &\approx&
T\,\left(\frac{10\pi^2}{3 H_{100}^2}\right)^2\,
\left[\,\int_{-\infty}^\infty df\, 
\frac{\gamma^2(|f|)}{f^6 P_1(|f)P_2(|f|)}\right]^{-1}\,.
\label{e:sigma2_continuous_final}
\end{eqnarray}
In the sense that $\widetilde Q(f)$ maximizes $\mu_Y/\sigma_Y$, it is
the {\em optimal filter\,} for the cross-correlation $Y$. 
The signal-to-noise ratio $\rho_Y\equiv Y/\sigma_Y$ has expected value
\begin{eqnarray}
\langle\rho_Y\rangle &=& \frac{\mu_Y}{\sigma_Y}
\label{e:SNR}\\ 
&\approx&
\frac{3H_0^2}{10\pi^2}\,
\Omega_0\,\sqrt{T}\, 
\left[\int_{-\infty}^\infty df\ \frac{\gamma^2(|f|)}{f^6 P_1(|f|) P_2(|f|)}
\right]^{1/2}\,,
\nonumber
\end{eqnarray}
which grows with the square-root of the observation time $T$, and
inversely with the product of the {\em amplitude\,} noise spectral
densities of the two detectors. In order of magnitude,
Eq.~\ref{e:SNR} indicates that the upper limit we can place on
$\Omega_0\,h_{100}^2$ by cross-correlation is smaller (i.e., more
constraining) than that obtainable from one detector by a factor
of $\gamma_{\rm rms}\sqrt{T \Delta_{\rm BW}}$, where $\Delta_{\rm BW}$ 
is the bandwidth over which the integrand of Eq.~\ref{e:SNR} is
significant (roughly the width of the peak of $1/f^3 P_i(f)$), 
and $\gamma_{\rm rms}$ is the rms value of $\gamma (f)$ over that
bandwidth. For the LHO-LLO correlations in this analysis, $T\sim
2\times 10^5$~sec, $\Delta_{\rm BW} \sim 100$~Hz, and $\gamma_{\rm rms}
\sim 0.1$, so we expect to be able to set a limit that is a factor
of several hundred below the individual detectors' strain
noise\footnote{More precisely, if the two detectors have unequal
strain sensitivities, the cross-correlation limit will be a factor
of $\gamma_{\rm rms}\sqrt{T \Delta_{\rm BW}}$ below the geometric mean of
the two noise spectral densities.}, or $\Omega_0\,h_{100}^2 \sim
10$ as shown in Fig.~\ref{fig:sensitivityPlots}.

\section{ANALYSIS OF LIGO DATA}

\subsection{Data analysis pipeline}
\label{sec:pipeline}

\begin{figure}[htbp!]
  \includegraphics[width=3.5in,angle=0]{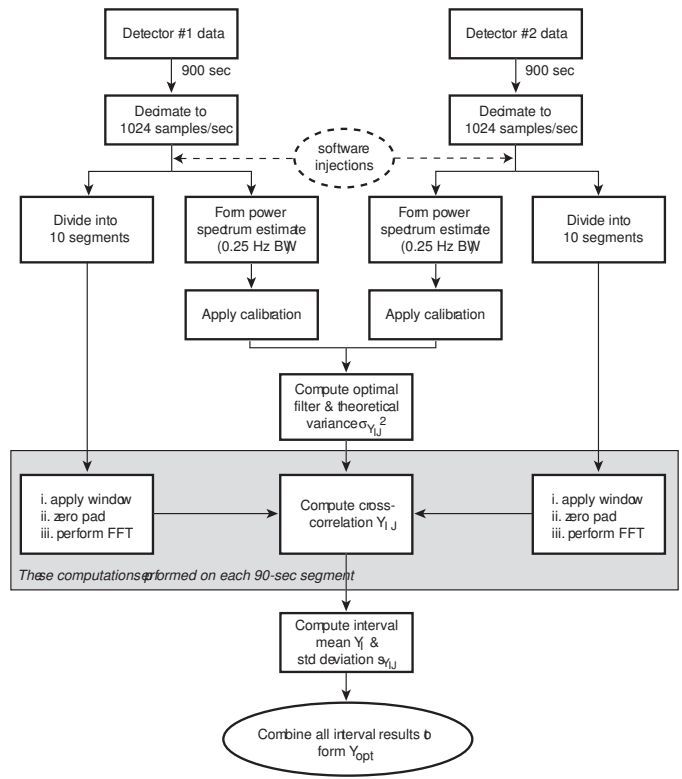}
  \caption{Data analysis flow diagram for the stochastic search.
  The raw detector signal (i.e., the uncalibrated differential arm
  error signal) is fed into the pipeline in 900-sec long intervals.
  Simulated stochastic background signals
  can be injected near the beginning of each data path, allowing us to
  test the data analysis routines in the presence of known correlations.}
  \label{fig:dataPipeline}
\end{figure}

\begin{figure}[htbp!]
  \includegraphics[height=3.5in,angle=-90]{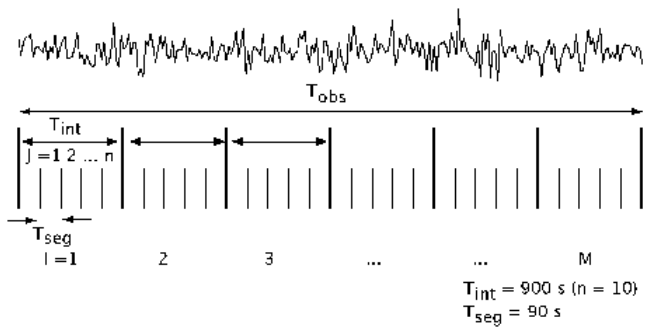}
  \caption{Time-series data from each interferometer is split
  into $M$ 900-sec intervals, which are further
  subdivided into $n=10$ 90-sec data segments.
  Cross-correlation values $Y_{IJ}$ are calculated
  for each 90-sec segment; theoretical variances $\sigma^2_{Y_{IJ}}$
  are calculated for each 900-sec interval.
  Here $I=1,2,\cdots, M$ labels the different intervals, and
  $J=1,2,\cdots, n$ labels the individual segments within
  each interval.}
  \label{fig:data_partitions}
\end{figure}

A flow diagram of the data analysis pipeline is shown in
Fig.~\ref{fig:dataPipeline} \cite{medusa}.
We perform the analysis in the frequency domain, where it is more
convenient to construct and apply the optimal filter. Since the
data are discretely-sampled, we use discrete Fourier transforms
and sums over frequency bins rather than integrals. The data
$r_i[k]$ are the raw (uncalibrated) detector outputs at discrete
times $t_k\equiv k\,\delta t$:
\begin{equation}
r_i[k]\equiv r_i(t_k)\,,
\label{e:r_i[k]}
\end{equation}
where $k=0,1,2,\cdots$, $\delta t$ is the sampling period, and $i$
labels the detector. We decimate the data to a sampling rate of
$(\delta t)^{-1} = 1024$~Hz (from 16384~Hz), since the higher frequencies
make a negligible contribution to the cross-correlation. The
decimation is performed with a finite impulse response filter of
length 320, and cut-off frequency 512~Hz. The data are split into
{\em intervals} (labeled by index $I$) and {\em segments} (labeled
by index $J$) within each interval to deal with detector
nonstationarity and to produce sets of cross-correlation values
$Y_{IJ}$ for which empirical variances can be calculated;
see Fig.~\ref{fig:data_partitions}.
The time-series data corresponding to the $J$-th segment in
interval $I$ is denoted $r_{iIJ}[k]$, where $k=0,1,\cdots, N-1$
runs over the total number of samples in the segment.

A single optimal filter $\widetilde Q_I$ is calculated and applied
for each interval $I$, the duration of which should be long enough
to capture relatively narrow-band features in the power spectra,
yet short enough to account for significant non-stationary
detector noise. Based on observations of detector noise variation,
we chose an interval duration of $T_{\rm int} = 900$~sec. The
segment duration should be much greater than the light travel time
between the two detectors, yet short enough to yield a sufficient
number of cross-correlation measurements within each interval to
obtain an experimental estimate of the theoretical variance
$\sigma^2_{Y_{IJ}}$ of the cross correlation statistic $Y_{IJ}$.
We chose a segment duration of $T_{\rm seg} = 90$~sec, yielding
ten $Y_{IJ}$ values per interval.

To compute the segment cross-correlation values $Y_{IJ}$, the raw
decimated data $r_{iIJ}[k]$ are windowed in the time domain (see
Sec.~\ref{sec:window} for details), zero-padded to twice their
length (to avoid wrap-around problems \cite{NRC} when calculating
the cross-correlation statistic in the frequency domain), and
discrete Fourier-transformed. Explicitly, defining
\begin{equation}
g_{iIJ}[k]\equiv\ \left\{ \begin{array}{cl}
w_i[k]\,r_{iIJ}[k]  &    k = 0, \cdots, N-1 \\
0                   &    k = N, \cdots, 2N-1\,,
\end{array}
\right.
\label{e:g_iIJ}
\end{equation}
where $w_i[k]$ is the window function for the $i$-th
detector,%
\footnote{In general, one can use different window functions for
different detectors. However, for the S1 analysis, we took
$w_1[k]=w_2[k]$.}
the discrete Fourier transform is:
\begin{equation}
\widetilde{g}_{iIJ}[q]\equiv  \sum_{k=0}^{2N-1} \delta t\
g_{iIJ}[k] e^{-i2\pi k q/2N}\,,
\end{equation}
where $N=T_{\rm seg}/\delta t=92160$ is the number of data points
in a segment, and $q = 0,1,\cdots,2N-1$. The cross-spectrum
$\widetilde{g}^*_{1IJ}[q]\cdot\widetilde{g}_{2IJ}[q]$ is formed
and binned to the frequency resolution, $\delta\!f$, of the
optimal filter $\widetilde Q_I$:%
\footnote{As discussed below, $\delta\!f=0.25$~Hz yielding 
$n_b=45$ and $m=22$.} 
\begin{equation}
\mathcal{G}_{IJ}[\ell] \equiv 
\frac{1}{n_b}
\sum_{q=n_b \ell\!-m}^{n_b\ell+m}
\widetilde{g}^*_{1IJ}[q]\,\widetilde{g}_{2IJ}[q]\,,
\label{e:G_12IJ}
\end{equation}
where
$\ell_{\rm min}\le \ell\le \ell_{\rm max}$,
$n_b = 2\,T_{\rm seg} \delta\!f$ is the number of frequency
values being binned, and $m = (n_b-1)/2$.
The index $\ell$ labels the
discrete frequencies, $f_{\ell}\equiv \ell\,\delta\!f$. The
$\mathcal{G}_{IJ}[\ell]$ are computed for a range of $\ell$
that includes only the
frequency band that yields most of the expected signal-to-noise
ratio (e.g., 40 to 314~Hz for the LHO-LLO correlations), as
described in Sec.~\ref{sec:freqselection}. The cross-correlation
values are calculated as:
\begin{equation}
Y_{IJ} \equiv 2\,{\rm Re}\,\left[\,\sum_{\ell=\ell_{\rm
min}}^{\ell_{\rm max}} \delta\!f\ \widetilde Q_I[\ell]\,
\mathcal{G}_{IJ}[\ell]\,\right]\,. 
\label{e:Y_discrete_freq}
\end{equation}
Some of the frequency bins within the $\{\ell_{\rm min},\ell_{\rm
max}\}$ range are excluded from the above sum to avoid narrow-band
instrumental correlations, as described in
Sec.~\ref{sec:freqselection}. Except for the details of windowing,
binning, and band-limiting, Eq.~\ref{e:Y_discrete_freq} for
$Y_{IJ}$ is just a discrete-frequency approximation to
Eq.~\ref{e:Y_continuous_freq} for $Y$ for the continuous-frequency
data, with $df'\,\delta_T(f-f')$ approximated by a Kronecker delta
$\delta_{\ell\ell'}$ in discrete frequencies $f_\ell$ and
$f_{\ell'}$.%
\footnote{To make this correspondence with 
Eq.~\ref{e:Y_continuous_freq}, the factor of $2$ and real part 
in Eq.~\ref{e:Y_discrete_freq} are needed since we are summing 
only over {\em positive} frequencies, e.g., 40 to 314~Hz for 
the LHO-LLO correlation.
Basically, integrals over continuous frequency are replaced by 
sums over discrete frequency bins using the correspondence
$\int_{-\infty}^\infty\,df \rightarrow 
2\,{\rm Re}\,\sum_{\ell=\ell_{\rm min}}^{\ell_{\rm max}}\,\delta\!f$.}


In calculating the optimal filter, we estimate the strain noise
power spectra $P_{iI}$ for the interval $I$ using Welch's method:
449 periodograms are formed and averaged from 4096-point,
Hann-windowed data segments, overlapped by 50\%, giving a
frequency resolution $\delta f=0.25$~Hz. 
To calibrate the spectra in strain, we apply the calibration 
response function $\widetilde R_i(f)$ which converts the raw data 
to equivalent strain: 
$\widetilde s_i(f) = \widetilde R_i^{-1}(f) \widetilde r_i(f)$. 
The calibration lines described in Sec.~\ref{sec:instrument}
were measured once per 60 seconds; for each interval $I$, we 
apply the response function, $\widetilde R_{iI}$, corresponding 
to the middle 60 seconds of the interval.
The optimal filter $\widetilde Q_I$ for the case $\Omega_{\rm
gw}(f) \equiv\Omega_0 = {\rm const}$ is then constructed as:
\begin{equation}
\widetilde Q_I[\ell]\equiv{\cal N}_I \frac{\gamma[\ell]}
{|f_\ell|^3 (\widetilde R_{1I}[\ell]P_{1I}[\ell])^* (\widetilde
R_{2I}[\ell]P_{2I}[\ell])}\,, \label{e:optimal_discrete}
\end{equation}
where $\gamma[\ell]\equiv\gamma(f_\ell)$, and $\widetilde
R_{iI}[\ell]\equiv\widetilde R_{iI}(f_\ell)$.
By including the additional response function factors $\widetilde
R_{iI}$ in Eq.~\ref{e:optimal_discrete}, $\widetilde Q_I$ has the
appropriate units to act directly on the raw detector outputs in
the calculation of $Y_{IJ}$ (c.f.\ Eq.~\ref{e:Y_discrete_freq}).

The normalization factor ${\cal N}_I$ in
Eq.~\ref{e:optimal_discrete} takes into account the effect of
windowing \cite{windowing}. Choosing ${\cal N}_I$ so that the
theoretical mean of the cross-correlation $Y_{IJ}$ is equal to
$\Omega_0\,h_{100}^2\,T_{\rm seg}$ for all $I$, $J$ (as was done
for $Y$ in Sec.~\ref{sec:ccstatistic}),
we have:
\begin{eqnarray}
{\cal N}_I&=&
\frac{20\pi^2}{3H_{100}^2}\,
\frac{1}{\overline{w_1w_2}}\,\times
\label{e:normalization_discrete}\\
&\times&\,
\left[2\,\,\sum_{\ell=\ell_{\rm min}}^{\ell_{\rm max}}\delta\!f\,
\frac{\gamma^2[\ell]}{f_\ell^6 P_{1I}[\ell] P_{2I}[\ell]}
\,\right]^{-1}\,,
\nonumber\\
\sigma_{Y_{IJ}}^2&=&
T_{\rm seg}\, \left(\frac{10 \pi^2}{3 H_{100}^2}\right)^2\,
\frac{\overline{w_1^2 w_2^2}}{(\overline{w_1w_2})^2}\,\times
\label{e:sigma2_discrete_final}\\
&\times&\,
\left[\, 2\,\sum_{\ell=\ell_{\rm min}}^{\ell_{\rm max}} \delta\!f\,
\frac{\gamma^2[\ell]}{f_\ell^6 P_{1I}[\ell] P_{2I}[\ell]}
\,\right]^{-1}\,,
\nonumber
\end{eqnarray}
where
\begin{eqnarray}
\overline{w_1w_2}&\equiv&
\frac{1}{N}\sum_{k=0}^{N-1}w_1[k]w_2[k]\,,
\label{e:window_fac1}\\
\overline{w_1^2 w_2^2}&\equiv&
\frac{1}{N}\sum_{k=0}^{N-1}w_1^2[k]w_2^2[k]\,,
\label{e:window_fac2}
\end{eqnarray}
provided the windowing is sufficient to prevent significant leakage
of power across the frequency band (see Sec.~\ref{sec:window}
and \cite{windowing} for more details).
Note that the theoretical variance $\sigma^2_{Y_{IJ}}$ depends
only on the interval $I$, since the cross-correlations $Y_{IJ}$
have the same statistical properties for each segment $J$ in $I$.

For each interval $I$, we calculate the mean, $Y_I$, and (sample)
standard deviation, $s_{Y_{IJ}}$, of the 10 cross-correlation
values $Y_{IJ}$:
\begin{eqnarray}
Y_{I}&\equiv& {1\over 10}\sum_{J=1}^{10}Y_{IJ}\,,
\label{e:Y_I}\\
s_{Y_{IJ}} &\equiv& \sqrt{ {1\over 9}\sum_{J=1}^{10}(Y_{IJ}-Y_{I})^2 }\,.
\label{e:s_I}
\end{eqnarray}
We also form a {\em weighted} average, $Y_{\rm opt}$, of the $Y_I$
over the whole run:
\begin{equation}
Y_{\rm opt}\equiv\frac{\sum_I\sigma_{Y_{IJ}}^{-2}\,Y_{I}}
{\sum_I\sigma_{Y_{IJ}}^{-2}}\,.
\label{e:Yoptimal}
\end{equation}
The statistic $Y_{\rm opt}$ maximizes the expected
signal-to-noise ratio for a stochastic signal, allowing for
non-stationary
detector noise from one 900-sec interval $I$ to the next
\cite{allenromano}.  Dividing $Y_{\rm opt}$ by the time $T_{\rm seg}$
over which an individual cross-correlation measurement is made gives,
in the absence of cross-correlated detector noise, an unbiased
estimate of the stochastic background level:%
\footnote{We use a hat $\widehat{\ }$ to indicate an estimate of
the actual (unknown) value of a quantity.}
$\widehat\Omega_0\,h_{100}^2 = Y_{\rm opt}/T_{\rm seg}$.

Finally, in Sec.~\ref{sec:fdomain} we will be interested in
the spectral properties of $Y_{IJ}$, $Y_{I}$, and $Y_{\rm opt}$.
Thus, for later reference, we define:
\begin{eqnarray}
\widetilde Y_{IJ}[\ell]&\equiv& \widetilde
Q_I[\ell]\,{\cal G}_{IJ}[\ell]\,,
\label{e:Y_IJ[l]}\\
\widetilde Y_I[\ell]&\equiv&
{1\over 10}\sum_{J=1}^{10}\widetilde Y_{IJ}[\ell]\,,
\label{e:Y_I[l]}\\
\widetilde Y_{\rm opt}[\ell]&\equiv&
\frac{\sum_I\sigma_{Y_{IJ}}^{-2}\,\widetilde Y_{I}[\ell]}
{\sum_I\sigma_{Y_{IJ}}^{-2}}\,.
\label{e:Y_opt[l]}
\end{eqnarray}
Note that $2\,{\rm Re}\,\sum_{\ell=\ell_{\rm min}}^{\ell_{\rm max}}
\delta\!f\ \cdot\ $ of the above quantities equal 
$Y_{IJ}$, $Y_I$, and $Y_{\rm opt}$, respectively. 

\subsection{Windowing}
\label{sec:window}

In taking the discrete Fourier transform of the raw 90-sec data
segments, care must be taken to limit the spectral leakage of large,
low-frequency components into the sensitive band. In general, some
combination of high-pass filtering in the time domain, and windowing
prior to the Fourier transform can be used to deal with spectral
leakage.  In this analysis we have found it sufficient to apply an
appropriate window to the data.

Examining the dynamic range of the data helps establish the
allowed leakage. Figure \ref{fig:sensitivityPlots} shows that the
lowest instrument noise around 60~Hz is approximately
$10^{-19}/\sqrt{{\rm Hz}}$ (for L1). While not shown in this plot,
the rms level of the raw data corresponds to a
strain of order $10^{-16}$, and is due to fluctuations in the
$10-30$~Hz band. Leakage of these low-frequency components must
be at least below the sensitive band noise level; e.g., leakage
must be below $10^{-3}$ for a $~30$~Hz offset. A tighter
constraint on the leakage comes when considering that these
low-frequency components may be correlated between the two
detectors, as they surely will be at some frequencies for the two
interferometers at LHO, due to the common seismic environment. In
this case the leakage should be below the predicted stochastic
background sensitivity level, which is approximately 2.5 orders
of magnitude below the individual detector noise levels for the
LHO H1-H2 case. Thus, the leakage should be below $3 \times
10^{-6}$ for a $~30$~Hz offset.

On the other hand, we prefer not to use a window that has an
average value significantly less than unity (and correspondingly
low leakage, such as a Hann window), because it will effectively
reduce the amount of data contributing to the cross-correlation.
Provided that the windowing is sufficient to prevent
significant leakage of power across the frequency range, the
net effect is to multiply the expected value of the signal-to-noise
ratio by $\overline{w_1w_2}/\sqrt{\overline{w_1^2 w_2^2}}$
(c.f.\ Eqs.~\ref{e:normalization_discrete}, \ref{e:sigma2_discrete_final}).

For example, when $w_1$ and $w_2$ are both Hann windows, this
factor is equal to $\sqrt{18/35}\approx 0.717$, which is
equivalent to reducing the data set length by a factor of 2. In
principle one should be able to use overlapping data segments to
avoid this effective loss of data, as in Welch's power spectrum
estimation method. In this case, the calculations for the expected
mean and variance of the cross-correlations would have
to take into account the statistical interdependence of the
overlapping data.

\begin{figure}[htbp!]
  \includegraphics[width=3.5in,angle=0]{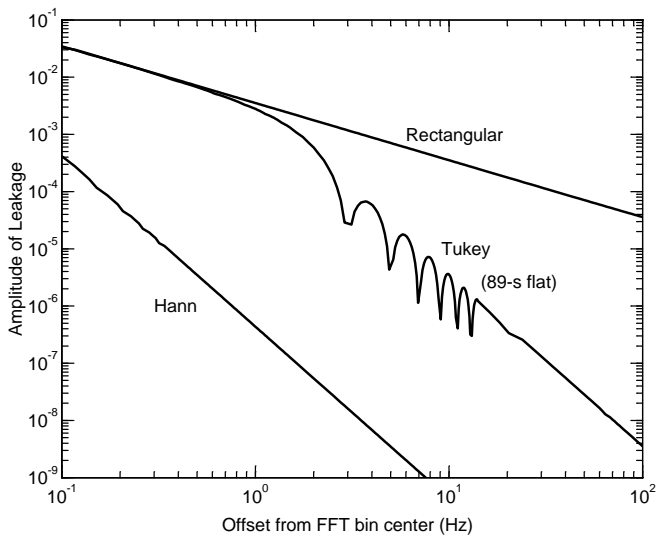}
  \caption{Leakage function for a rectangular window, a standard
    Hann window of width 90-sec, and a Tukey window
    consisting of a 1-sec Hann window with an 89-sec flat
    section spliced into the middle. The curves show the envelope of
    the leakage functions, with a varying frequency resolution, so the
    zeros of the functions are not seen.}
  \label{fig:flattopLeakage}
\end{figure}

Instead, we have used a Tukey window \cite{dspbook}, which is
essentially a Hann window split in half, with a constant section
of all 1's in the middle.
%
%
We can choose the length of the Hann portion of the window to
provide sufficiently low leakage, yet maintain a unity value over
most of the window. Figure \ref{fig:flattopLeakage} shows the
leakage function of the Tukey window that we use
(a 1-sec Hann window with an 89-sec flat section spliced into the
middle), and compares it to Hann and rectangular windows. The
Tukey window leakage is less than $10^{-7}$ for all frequencies
greater than $~35$ Hz away from the FFT bin center.
This is 4 orders of magnitude better than what is
needed for the LHO-LLO correlations and a factor of 30 better
suppression than needed for the H1-H2 correlation.

To explicitly verify that the Tukey window behaved as expected,
we re-analyzed the H1-H2 data with a pure Hann window
(see also Sec.~\ref{sec:H1H2correlation}).
The result of this re-analysis, properly scaled to take into
account the effective reduction in observation time, was, within
error, the same as the original analysis with a Tukey window.
Since the H1-H2 correlation is the most prone of all correlations
to spectral leakage (due to the likelihood of cross-correlated
low-frequency noise components), the lack of a significant
difference between the pure Hann and Tukey window analyses
provided additional support for the use of the Tukey window.

\subsection{Frequency band selection and discrete frequency elimination}
\label{sec:freqselection}

\begin{figure}[htbp!]
  \includegraphics[width=3.5in,angle=0]{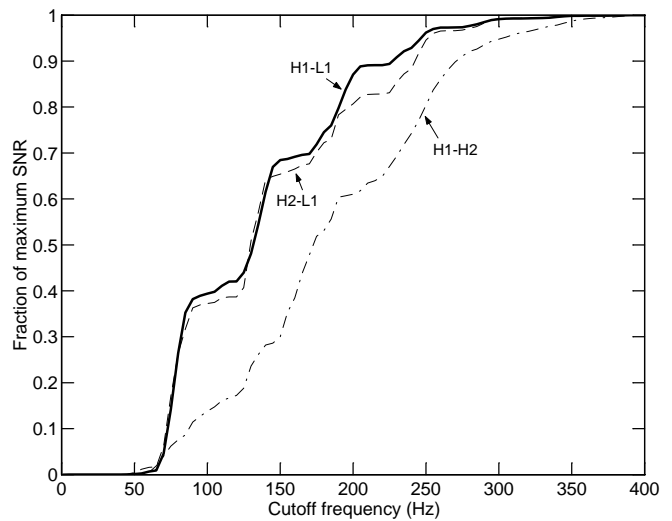}
  \caption{Curves show the fraction of maximum expected 
  signal-to-noise ratio as a
  function of cut-off frequency, for the three interferometer
  pairs. The curves were made by numerically integrating
  Eq.~\ref{e:SNR} from a few Hz up to the variable cut-off
  frequency, using the strain sensitivity spectra shown in
  Fig.~\ref{fig:sensitivityPlots}.}
  \label{fig:snrvscutoff}
\end{figure}

In computing the discrete cross-correlation integral, we are free
to restrict the sum to a chosen frequency region or regions; in
this way the variance can be reduced (e.g., by excluding low
frequencies where the detector power spectra are large and
relatively less stationary), while still retaining most of the
signal. We choose the frequency ranges by determining the band
that contributes most of the expected signal-to-noise ratio,
according to Eq.~\ref{e:SNR}. Using the strain power spectra
shown in Fig.~\ref{fig:sensitivityPlots}, we compute the
signal-to-noise ratio integral of Eq.~\ref{e:SNR} from a very low
frequency (a few Hz) up to a variable cut-off frequency, and plot
the resulting signal-to-noise ratio versus cut-off frequency
(Fig.~\ref{fig:snrvscutoff}). For each interferometer pair, the
lower band edge is chosen to be 40~Hz, while the upper band edge
choices are 314~Hz for LHO-LLO correlations (where there is a
zero in the overlap reduction function), and 300~Hz for H1-H2
correlations (chosen to exclude $\sim340$~Hz resonances in the
test mass mechanical suspensions, which were not well-resolved
in the power spectra).

Within the $40-314$ (300)~Hz band, discrete frequency bins at which
there are known or potential instrumental correlations due to
common periodic sources are eliminated from the cross-correlation
sum. For example, a significant feature in all interferometer
outputs is a set of spectral lines extending out to beyond 2~kHz,
corresponding to the 60~Hz power line and its harmonics ($n\cdot
60$~Hz lines). Since these lines obviously have a common
source---the mains power supplying the instrumentation---they are
potentially correlated between detectors. To avoid including any
such correlation in the analysis, we eliminate the $n\cdot 60$~Hz
frequency bins from the sum in Eq.~\ref{e:Y_discrete_freq}.

Another common periodic signal arises from the data acquisition
timing systems in the detectors. The absolute timing and
synchronization of the data acquisition systems
between detectors is based on 1 pulse-per-second
signals produced by Global Positioning System (GPS) receivers at
each site. In each detector, data samples are stored temporarily
in $1/16$~sec buffers, prior to being collected and written to
disk. The process, through mechanisms not yet established,
results in some power at 16~Hz and harmonics in the detectors'
output data channels. These signals are extremely narrow-band
and, due to the stability and common source of the GPS-derived
timing, can be correlated between detectors. To avoid including
any of these narrow-band correlations, we eliminate the $n\cdot
16$~Hz frequency bins from the sum in Eq.~\ref{e:Y_discrete_freq}.

Finally, there may be additional correlated narrow-band features
due to highly stable clocks or oscillators that are common
components among the detectors (e.g., computer monitors can have
very stable sync rates, typically at 70~Hz). To describe how we
avoid such features, we first present a quantitative analysis of
the effect of coherent spectral lines on our cross-correlation
measurement. We begin by following the treatment of correlated
detector noise given in Sec.~V.E of Ref.~\cite{allenromano}. The
contribution of cross-correlated detector noise to the
cross-correlation $Y$ will be small compared to the
intrinsic measurement noise if
\begin{equation}
\left|\frac{T}{2}\int_{-\infty}^{\infty}
df\ P_{12}(|f|)\widetilde Q(f)\right|
\ll\sigma_Y\,, 
\label{e:condition1}
\end{equation}
where $P_{12}(|f|)$ is the cross-power spectrum of the strain noise
($n_1$, $n_2$) in the two detectors, $T$ is the total observation
time, and $\sigma_Y$ is defined by Eq.~\ref{e:sigma2_continuous}.
Using Eq.~\ref{e:optimal_continuous}, this condition becomes
\begin{equation}
{\cal N} \left|\int_{-\infty}^\infty df\ \frac{P_{12}(|f|)
\gamma(|f|)}{|f|^3 P_1(|f|)P_2(|f|)}\right| \ll 
\frac{2\sigma_Y}{T}\,, 
\label{e:condition2}
\end{equation}
or, equivalently,
\begin{equation}
\frac{3 H_{100}^2}{5 \pi^2}
\left|\int_{-\infty}^\infty df\ \frac{P_{12}(|f|)
\gamma(|f|)}{|f|^3 P_1(|f|)P_2(|f|)}\right| \ll 
2\sigma_Y^{-1}\,, 
\label{e:condition3}
\end{equation}
where Eqs.~\ref{e:normalization_continuous}, 
\ref{e:sigma2_continuous_final} 
were used to eliminate ${\cal N}$ in terms of $\sigma_Y^2$:
\begin{equation}
{\cal N}=\frac{3 H_{100}^2}{5\pi^2}\,\frac{\sigma_Y^2}{T}\,.
\end{equation}
Now consider the presence of a correlated periodic signal, such
that the cross-spectrum $P_{12}(f)$ is significant only at a
single (positive) discrete frequency, $f_L$. For this component to 
have a small effect, the above condition becomes:
\begin{equation}
\frac{3 H_{100}^2}{5 \pi^2} \left|\Delta f
\frac{P_{12}(f_L)\gamma(f_L)}
{f_L^3 P_1(f_L)P_2(f_L)}\right| \ll
\sigma_Y^{-1}\,,
\label{e:condition4}
\end{equation}
where $\Delta f$ is the frequency resolution of the discrete 
Fourier transform used to approximate the frequency integrals, and
the left-hand-side of Eq.~\ref{e:condition4} can be expressed in
terms of the coherence function $\Gamma_{12}(f)$, which is
essentially a normalized cross-spectrum, defined as \cite{bendat}:
\begin{equation}
\Gamma_{12}(f) \equiv \frac{| P_{12}(f) |^2} {P_1(f) P_2(f)}\ .
\label{e:coherencedef}
\end{equation}
The condition on the coherence at $f_L$ is thus
\begin{equation}
[\Gamma_{12}(f_L)]^{1/2} \ll 
\frac{\sigma_Y^{-1}}{\Delta f}\,
\frac{5\pi^2}{3 H_{100}^2}\,
\frac{\sqrt{P_1(f_L) P_2(f_L)}}
{|f_L^{-3}\gamma(f_L)|}\,.
\label{e:condition5}
\end{equation}

Since $\sigma_Y$ increases as $T^{1/2}$, the limit on the coherence
$\Gamma_{12}(f_L)$ becomes smaller as $1/T$. To show how this
condition applies to the S1 data, we estimate the factors in
Eq.~\ref{e:condition5} for the H2-L1 pair, focusing on the band
$100-150$~Hz. We assume any correlated spectral line is weak enough
that it does not appear in the power spectrum estimates used to
construct the optimal filter.  Noting that the combination $(3
H_{100}^2/10\pi^2)f^{-3}$ is just the power spectrum of gravitational
waves $S_{\rm gw}(f)$ with $\Omega_0\,h_{100}^2 = 1$ (c.f.\
Eq.~\ref{e:gwpsd}), we can evaluate the right-hand-side of
Eq.~\ref{e:condition5} by estimating the ratios $[P_i/S_{\rm
gw}]^{1/2}$ from Fig.~\ref{fig:sensitivityPlots} for
$\Omega_0\,h_{100}^2=1$.  Within the band
$100-150$~Hz, this gives: $(P_1 P_2)^{1/2}/S_{\rm gw} \gtrsim
2500$. The overlap reduction function in this band is $|\gamma|
\lesssim 0.05$. The appropriate frequency resolution $\Delta f$ is
that corresponding to the 90-sec segment discrete Fourier transforms,
so $\Delta f = 0.011$~Hz. As described later in Sec.~\ref{sec:errors}, 
we calculate a statistical error, $\sigma_{\Omega}$, associated with 
the stochastic background estimate $Y_{\rm opt}/T_{\rm seg}$.  
Under the implicit assumption made in
Eq.~\ref{e:condition5} that the detector noise is stationary, one can
show that $\sigma_Y = T\,\sigma_{\Omega}$.  Finally, referring to
Table~\ref{tab:errorestimates} for an estimate of $\sigma_{\Omega}$,
and using the total H2-L1 observation time of 51~hours, we obtain
$\sigma_Y \approx 2.8 \times 10^6$~sec. Thus, the condition of
Eq.~\ref{e:condition5} becomes: $[\Gamma_{12}(f_L)]^{1/2} \ll
1$.

Using this example estimate as a guide, specific lines are
rejected by calculating the coherence function between detector
pairs for the full sets of analyzed S1 data, and eliminating any
frequency bins at which $\Gamma_{12}(f_L) \ge 10^{-2}$. 
The coherence functions are calculated with a frequency resolution 
of 0.033~Hz, and approximately 20,000 (35,000) averages for the
LHO-LLO (LHO-LHO) pairs, corresponding to statistical uncertainty
levels $\sigma_\Gamma\equiv 1/N_{\rm avg}$ of approximately 
$5\times 10^{-5}$ ($3\times 10^{-5}$). The exclusion threshold 
thus corresponds to a cut on the coherence data of order 
$100\,\sigma_{\Gamma}$.

\begin{figure}[htbp!]
  \includegraphics[width=3.75in,angle=0]{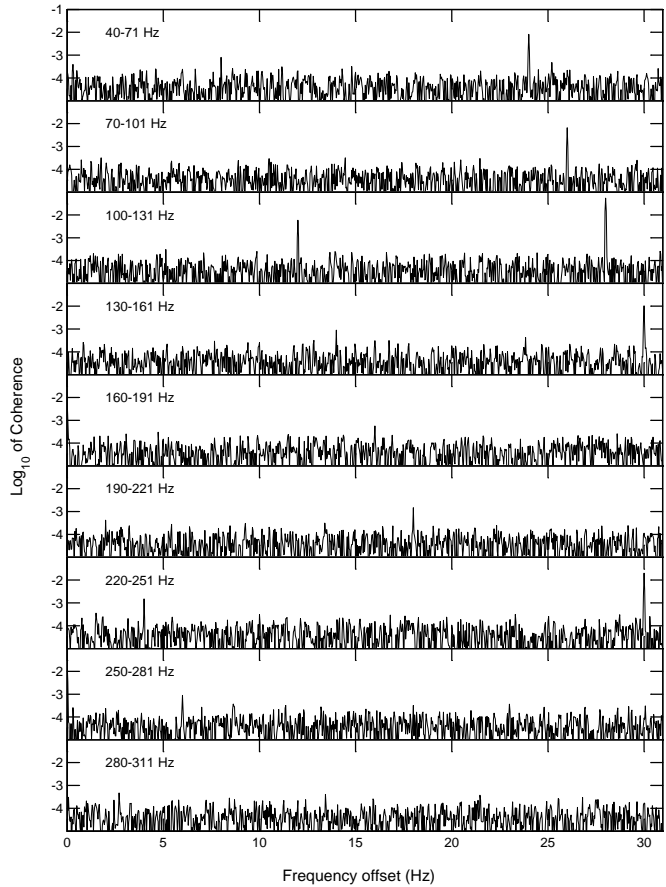}
  \caption{Coherence between the H2 and L1 detector outputs during S1.
  The coherence is calculated with a frequency resolution of 0.033~Hz
  and $N_{\rm avg} \approx 20,000$ periodogram averages (50\% overlap);
  Hann windows are used in the Fourier transforms. There are significant
  peaks at harmonics of 16~Hz (data acquisition buffer rate) and at
  250~Hz (unknown origin). These frequencies are all excluded from
  the cross-correlation sum. The broadband coherence level corresponds
  to the expected statistical uncertainty level of
  $1/N_{\rm avg} \approx 5 \times 10^{-5}$.}
  \label{fig:h2l1coherence}
\end{figure}

\begin{figure}[htbp!]
  \includegraphics[width=3.75in,angle=0]{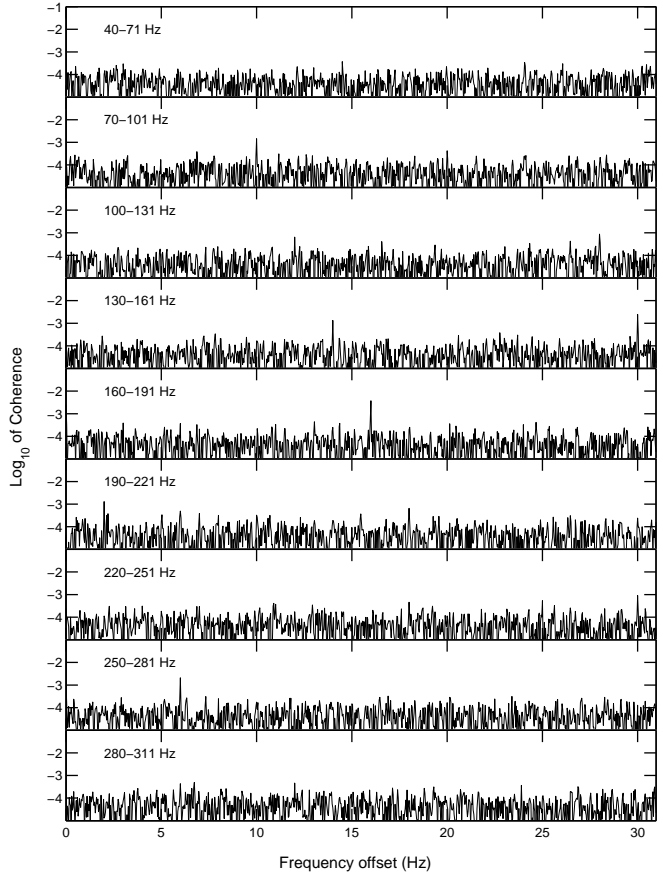}
  \caption{Coherence between the H1 and L1 detector outputs during
  S1, calculated as described in the caption of Fig.~\ref{fig:h2l1coherence}.
  There are significant peaks at harmonics of 16~Hz (data acquisition
  buffer rate). These frequencies are excluded from the cross-correlation sum.
  The broadband coherence level corresponds to the expected statistical
  uncertainty level of $1/N_{\rm avg} \approx 5 \times 10^{-5}$.}
  \label{fig:h1l1coherence}
\end{figure}

For the H2-L1 pair, this procedure results in eliminating the
250~Hz frequency bin, whose coherence level was about 0.02; the
H2-L1 coherence function over the analysis band is shown in
Fig.~\ref{fig:h2l1coherence}. For H1-H2, the bins at 168.25~Hz and
168.5~Hz were eliminated, where the coherence was also about 0.02
(see Fig.~\ref{fig:h1h2coherence}). The sources of these lines are
unknown. For H1-L1, no additional frequencies were removed by the
coherence threshold (see Fig.~\ref{fig:h1l1coherence}).

%

%

\begin{figure}[htbp!]
  \includegraphics[width=3.5in,angle=0]{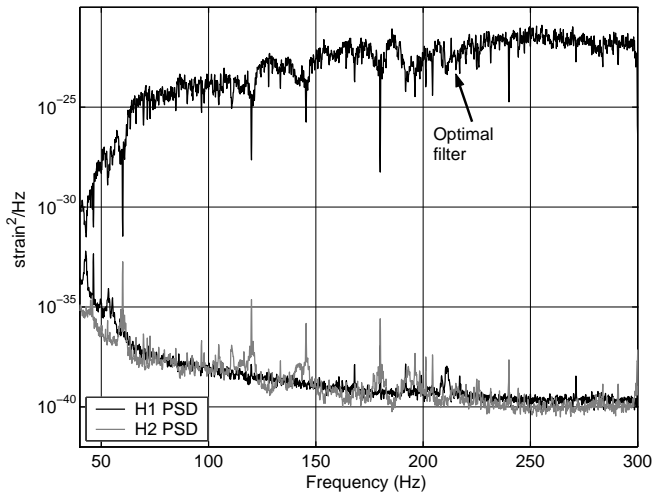}
  \caption{Power spectral densities and optimal filter for the
  H1-H2 detector pair, using the sensitivities shown in
  Fig.~\ref{fig:sensitivityPlots}. A scale factor has been applied
  to the optimal filter for display purposes. Note that spectral lines
  at 60~Hz and harmonics produce corresponding deep `notches' in the
  optimal filter.}
  \label{fig:optimalfiltplot}
\end{figure}

It is worth noting that correlations at the $n\cdot 60$~Hz lines
are suppressed even without explicitly eliminating these
frequency bins from the sum. This is because these frequencies
have a high signal-to-noise ratio in the power spectrum estimates,
and thus they have relatively small values in the optimal filter.
The optimal filter thus tends to suppress spectral lines that
show up in the power spectra. This effect is illustrated in
Fig.~\ref{fig:optimalfiltplot}, and is essentially the result of
having four powers of $\widetilde s_i(f)$ in the denominator of the
integrand of the cross-correlation, but only two powers
in the numerator. Nonetheless, we chose to remove the $n\cdot
60$~Hz bins from the cross-correlation sum for robustness, and
as good practice for future analyses, where improvements in the
electronics instrumentation may reduce the power line coupling
such that the optimal filter suppression is insufficient.

Such optimal filter suppression does not occur, however, for the
16~Hz line and its harmonics, and the additional 168.25, 168.5,
250~Hz lines; these lines typically do not appear in the power
spectrum estimates, or do so only with a small signal-to-noise
ratio. These lines must be explicitly eliminated from the
cross-correlation sum.
These discrete frequency bins are all zeroed out in the optimal
filter, so that each excluded frequency removes 0.25~Hz of
bandwidth from the calculation.


\subsection{Results and interpretation}
\label{sec:results}

The primary goal of our analysis is to set an upper limit on the
strength of a stochastic gravitational wave background. The
cross-correlation measurement is, in principle, sensitive to a
combination of a stochastic gravitational background and
instrumental noise that is correlated between two detectors. In
order to place an upper limit on a gravitational wave background,
we must have confidence that instrumental correlations are not
playing a significant role. Gaining such confidence for the
correlation of the two LHO interferometers may be difficult, in
general, as they are both exposed to many of the same
environmental disturbances. In fact, for the S1 analysis, a strong
(negative) correlation {\em was\,} observed between the two
Hanford interferometers, thus preventing us from setting an upper
limit on $\Omega_0\,h_{100}^2$ using the H1-H2 pair results. The
correlated instrumental noise sources, relatively broadband
compared to the excised narrow-band features described in the
previous section, produced a significant H1-H2 cross-correlation 
(signal-to-noise ratio of $-8.8$); see Sec.~\ref{sec:H1H2correlation} 
for further discussion of the H1-H2 instrumental correlations.

On the other hand, for the widely separated (LHO-LLO)
interferometer pairs, there are only a few paths through which
instrumental correlations could arise. Narrow-band inter-site
correlations are seen, as described in the previous section, but
the described measures have been taken to exclude them from the
analysis. Seismic and acoustic noise in the several to tens of Hz
band have characteristic coherence lengths of tens of meters or
less, compared to the 3000~km LHO-LLO separation, and pose little
problem. Globally correlated magnetic field fluctuations have been
identified in the past as the most likely candidate capable of
producing broadband correlated noise in the widely separated
detectors \cite{chris}. An order-of-magnitude analysis of this
effect was made in Ref.~\cite{allenromano}, concluding that
correlated field fluctuations during magnetically noisy periods
(such as during thunderstorms) could produce a LHO-LLO correlated
strain signal corresponding to a stochastic gravitational
background $\Omega_0\,h_{100}^2$ of order $10^{-8}$. These estimates
evaluated the forces produced on the test masses by the correlated
magnetic fields, via magnets that are bonded to the test masses to
provide position and orientation control.\footnote{The actual
limit on $\Omega_0\,h_{100}^2$ that appears in
Ref.~\cite{allenromano} is $10^{-7}$, since the authors assumed a
magnetic dipole moment of the test mass magnets that is a factor
of 10 higher than what is actually used.}
Direct tests made on the LIGO interferometers indicate that the
magnetic field coupling to the strain signal was generally much higher
during S1---up to $10^{2}$ times greater for a single
interferometer---than these force coupling estimates. Nonetheless, the
correspondingly modified estimate of the equivalent $\Omega$ due to
correlated magnetic fields is still 5 orders-of-magnitude below our
present sensitivity. Indeed, Figs.~\ref{fig:h2l1coherence} and
\ref{fig:h1l1coherence} show no evidence of any significant broadband
instrumental correlations in the S1 data. We thus set upper limits on
$\Omega_0\,h_{100}^2$ for both the H1-L1 and the H2-L1 pair results.

Accounting for the combination of a gravitational wave background
and instrumental correlations, we define an \emph{effective}
$\Omega$, $\Omega_{\rm eff}$, 
for which our measurement $Y_{\rm opt}/T_{\rm seg}$
provides an estimate:
\begin{equation}
\widehat\Omega_{\rm eff}\,h_{100}^2
\equiv 
Y_{\rm opt}/T_{\rm seg} 
= (\widehat\Omega_0 + \widehat\Omega_{\rm inst})\,h_{100}^2\,.
\label{e:hat_Omega_eff}
\end{equation}
Note that $\Omega_{\rm inst}$
(associated with instrumental correlations) may be either positive 
or negative, while $\Omega_0$ for the gravitational wave background
must be non-negative. We calculate a standard two-sided, frequentist 
90\% confidence interval on $\Omega_{\rm eff}\,h_{100}^2$ as follows:
\begin{equation}
  [\widehat\Omega_{\rm eff}\,h_{100}^2 - 
   1.65\,\widehat{\sigma}_{\Omega,{\rm tot}}\,,\ 
   \widehat\Omega_{\rm eff}\,h_{100}^2 + 
   1.65\,\widehat{\sigma}_{\Omega,{\rm tot}}]
  \label{e:confinterval}
\end{equation}
where $\widehat{\sigma}_{\Omega,{\rm tot}}$ is the total estimated
error of the cross-correlation measurement, as explained in 
Sec.~\ref{sec:errors}. 
In a frequentist interpretation, this means that if the
experiment were repeated many times, generating many values of
$\widehat\Omega_{\rm eff}\,h_{100}^2$ and $\widehat\sigma_{\Omega,{\rm
tot}}$, then the true value of $\Omega_{\rm eff}\,h_{100}^2$ is
expected to lie within 90 percent of these intervals. We establish
such a confidence interval for each detector pair.

For the H1-L1 and H2-L1 detector pairs, we are confident in
assuming that systematic broad-band instrumental cross-correlations 
are insignificant, so the measurement of $\widehat\Omega_{\rm
eff}\,h_{100}^2$ is used to establish an upper limit on a
stochastic gravitational wave background. Specifically, assuming
Gaussian statistics with fixed rms deviation, 
$\widehat\sigma_{\Omega,{\rm tot}}$, we set a standard 90\% 
confidence level upper limit on
$\Omega_0\,h_{100}^2$. Since the actual value of $\Omega_0$ must
be non-negative, we set the upper limit to
$1.28\,\widehat\sigma_{\Omega,{\rm tot}}$ if the measured value 
of $\widehat\Omega_{\rm eff}\,h_{100}^2$ is less than
zero.\footnote{We are assuming here that a negative value of
$\widehat\Omega_{\rm eff}\,h_{100}^2$ is due to random statistical
fluctuations in the detector outputs and not to systematic
instrumental correlations.}
Explicitly,
\begin{equation}
\Omega_0\,h_{100}^2 \le {\rm max} 
\{\widehat\Omega_{\rm eff}\,h_{100}^2\,,\,0\} + 
1.28\,\widehat\sigma_{\Omega,{\rm tot}}\,.
\label{e:confupperlimit}
\end{equation}

Table~\ref{tab:production_results} summarizes the results obtained
in applying the cross-correlation analysis to the LIGO S1 data.
The most constraining (i.e., smallest) upper limit on a
gravitational wave stochastic background comes from the H2-L1
detector pair, giving $\Omega_0\,h_{100}^2 \le 23$. The significant
H1-H2 instrumental correlation is discussed
further in Sec.~\ref{sec:H1H2correlation}. The upper limits in
Table~\ref{tab:production_results} can be compared with the
expectations given in Fig.~\ref{fig:sensitivityPlots}, properly
scaling the latter for the actual observation times. The H2-L1
expected upper limit for 50 hours of data would be
$\Omega_0\,h_{100}^2 \le 14$. The difference between this number
and our result of 23 is due to the fact that, on average, the
detector strain sensitivities were poorer than those shown in
Fig.~\ref{fig:sensitivityPlots}.

\begin{table*}
\begin{ruledtabular}
\begin{tabular}{cccccccc}

Interferometer & $\widehat\Omega_{\rm eff}\,h_{100}^2$ &
$\widehat\Omega_{\rm eff}\,h_{100}^2/ \widehat\sigma_{\Omega,{\rm tot}}$ &
90\% confidence & 90\% confidence & $\chi_{\rm min}^2$ &
Frequency & Observation \\
pair & & & interval on $\Omega_{\rm eff}\,h_{100}^2$ & upper limit &
(per dof) & range & time  \\
\hline

H1-H2 & $-8.3$ & $-8.8$ &
$[-9.9 \pm 2.0\,,\,-6.8 \pm 1.4]$ &
$-$ &
4.9 & $40-300$~Hz & 100.25 hr \\

H1-L1 & 32 & 1.8 &
$[2.1 \pm .42\,,\,61 \pm 12]$ &
$\Omega_0\,h_{100}^2 \le 55 \pm 11$ &
0.96 & $40-314$~Hz & 64 hr \\

H2-L1 & 0.16 & 0.0094 &
$[-30 \pm 6.0\,,\,30 \pm 6.0]$ &
$\Omega_0\,h_{100}^2\le 23 \pm 4.6$ &
1.0 & $40-314$~Hz & 51.25 hr \\

\end{tabular}
\end{ruledtabular}
\caption{Measured 90\% confidence intervals and upper limits for
the three LIGO interferometer pairs, assuming $\Omega_{\rm
gw}(f)\equiv\Omega_0={\rm const}$ in the specified frequency
band. For all three pairs we compute a confidence interval
according to Eq.~\ref{e:confinterval}. For the LHO-LLO pairs, we
are confident in assuming the instrumental correlations are
insignificant, and an upper limit on a stochastic gravitational
background is computed according to Eq.~\ref{e:confupperlimit}.
Our established upper limit comes from the H2-L1 pair. The $\pm$
error bars given for the confidence intervals and upper limit
values derive from a $\pm 10\%$ uncertainty in the calibration
magnitude of each detector; see Sec.~\ref{sec:errors} and
Table~\ref{tab:errorestimates}.
The $\chi_{\rm min}^2$ per degree of freedom values are the result of
a frequency-domain comparison between the measured and theoretically
expected cross-correlations, described in Sec.~\ref{sec:fdomain}.}
\label{tab:production_results}
\end{table*}

In computing the Table~\ref{tab:production_results} numbers, some
data selection has been performed to remove times of higher than
average detector noise. Specifically, the theoretical variances
of all 900-sec intervals are calculated, and the sum of the
${\sigma_{Y_{IJ}}^{-2}}$ is computed. We then select the set of largest
${\sigma_{Y_{IJ}}^{-2}}$ (i.e., the most sensitive intervals) that
accumulate 95\% of the sum of all the weighting factors, and
include only these intervals in the
Table~\ref{tab:production_results} results. This selection
includes $75-85\%$ of the analyzed data, depending on the
detector pair. We also excluded an additional $\sim 10$
hours of H2 data near the beginning of S1 because of large
data acquisition system timing errors during this period.
The deficits between the
observation times given in Table~\ref{tab:production_results} and
the total S1 double-coincident times given in
Sec.~\ref{sec:instrument} are due to a combination of these and
other selections, spelled out in Table~\ref{tab:dataselection}.

\begin{table}
\begin{ruledtabular}
\begin{tabular}{cccc}

Selection criteria & H1-H2 & H1-L1 & H2-L1 \\
\hline
\textbf{\emph{A:}} All double- & 188 hr & 116 hr & 131 hr\\
coincidence data& $46\%$ & $28\%$ & $32\%$\\
\hline
\textbf{\emph{B: A}} plus $T_{\rm lock}> 900$-sec, & 134 hr & 75 hr & 81 hr\\
\& calibration monitored & $33\%$ & $18\%$ & $20\%$\\
\hline
\textbf{\emph{C: B}} plus valid GPS timing, & 119 hr & 75 hr & 66 hr\\
\& calibrations within bounds & $29\%$ & $18\%$ & $16\%$\\
\hline
\textbf{\emph{D: C}} plus quietest& 100 hr & 64 hr & 51 hr\\
intervals & $25\%$ & $16\%$ & $13\%$\\

\end{tabular}
\end{ruledtabular}
\caption{Summary of the selection criteria applied to the
double-coincidence data for S1. {\bf{\emph{A:}}} portion of the
408-hr S1 run having double-coincidence stretches greater than
600-sec; {\bf{\emph{B:}}} data portion satisfying criterion
\emph{A}, plus: data length is $\ge 900$-sec for the analysis
pipeline, {\emph{and}} the calibration monitoring was operational;
{\bf{\emph{C:}}} data portion satisfying criterion \emph{B},
plus: GPS timing is valid {\emph{and}} calibration data are
within bounds; {\bf{\emph{D:}}} data portion satisfying criterion
\emph{C}, plus: quietest data intervals that accumulate $95\%$ of
the sum of the weighting factors. This last data set was used in
the analysis pipeline.} \label{tab:dataselection}
\end{table}

\begin{figure}[htbp!]
  \includegraphics[width=3.5in,angle=0]{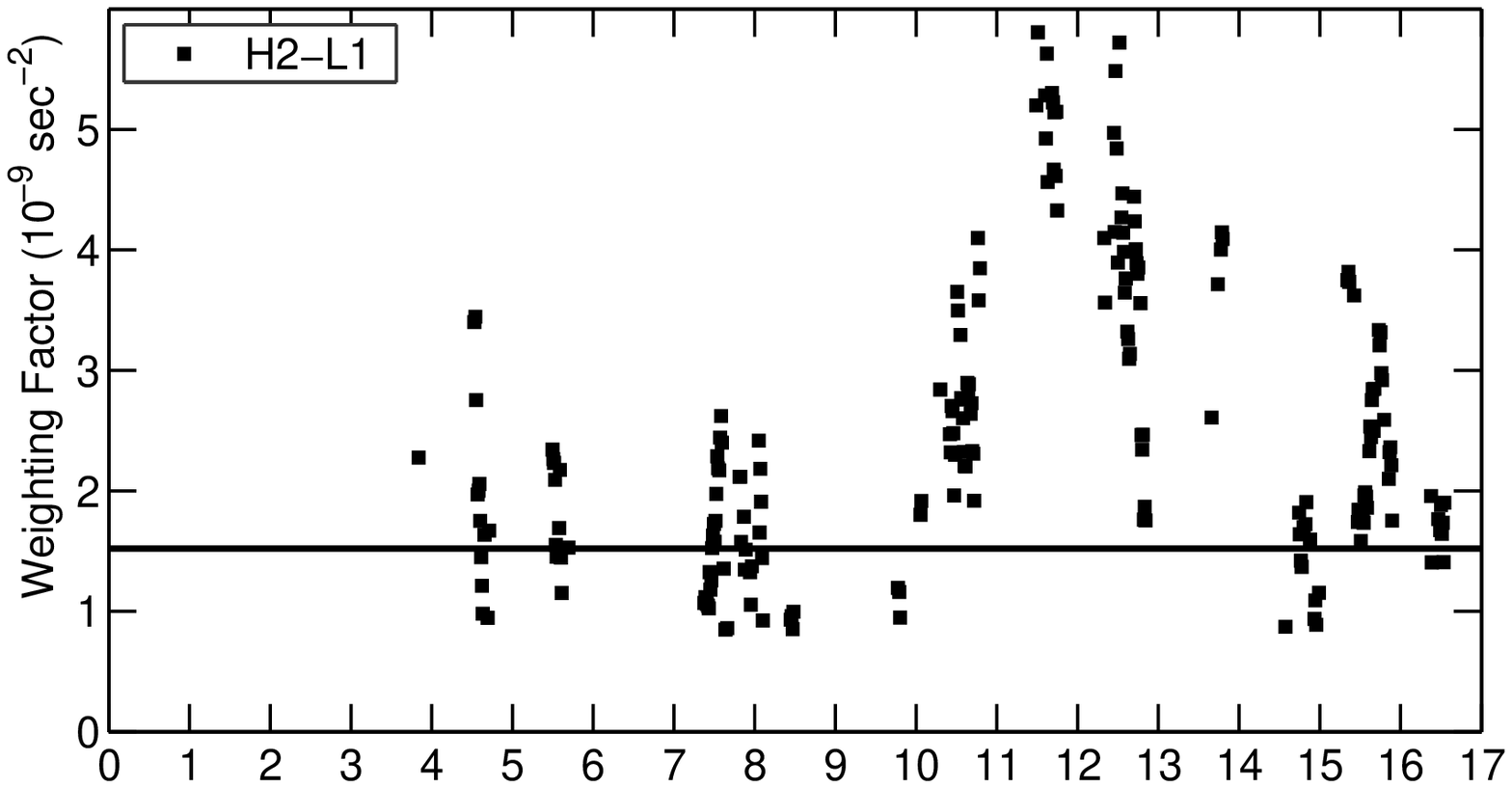}
  \includegraphics[width=3.5in,angle=0]{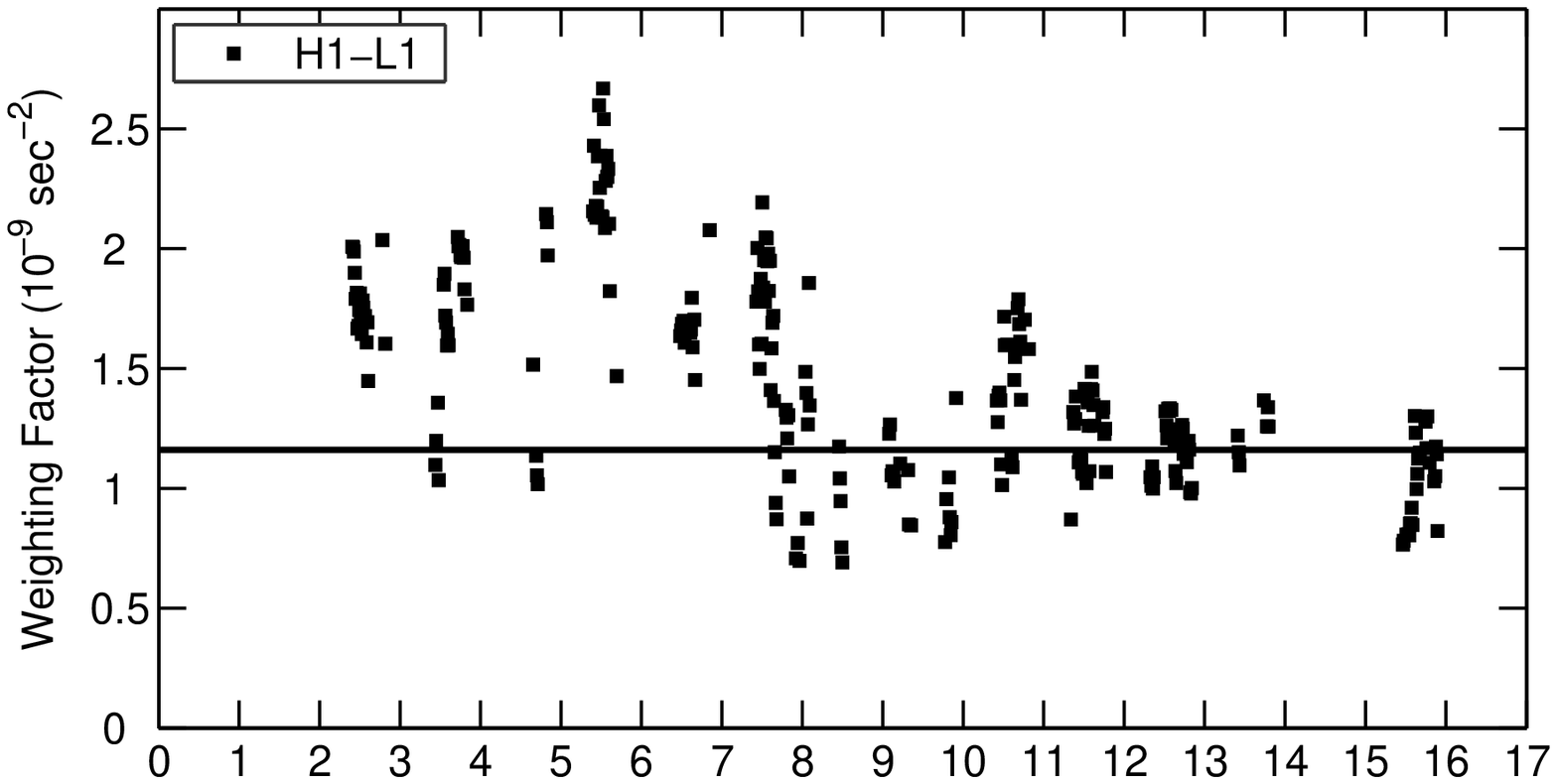}
  \includegraphics[width=3.5in,angle=0]{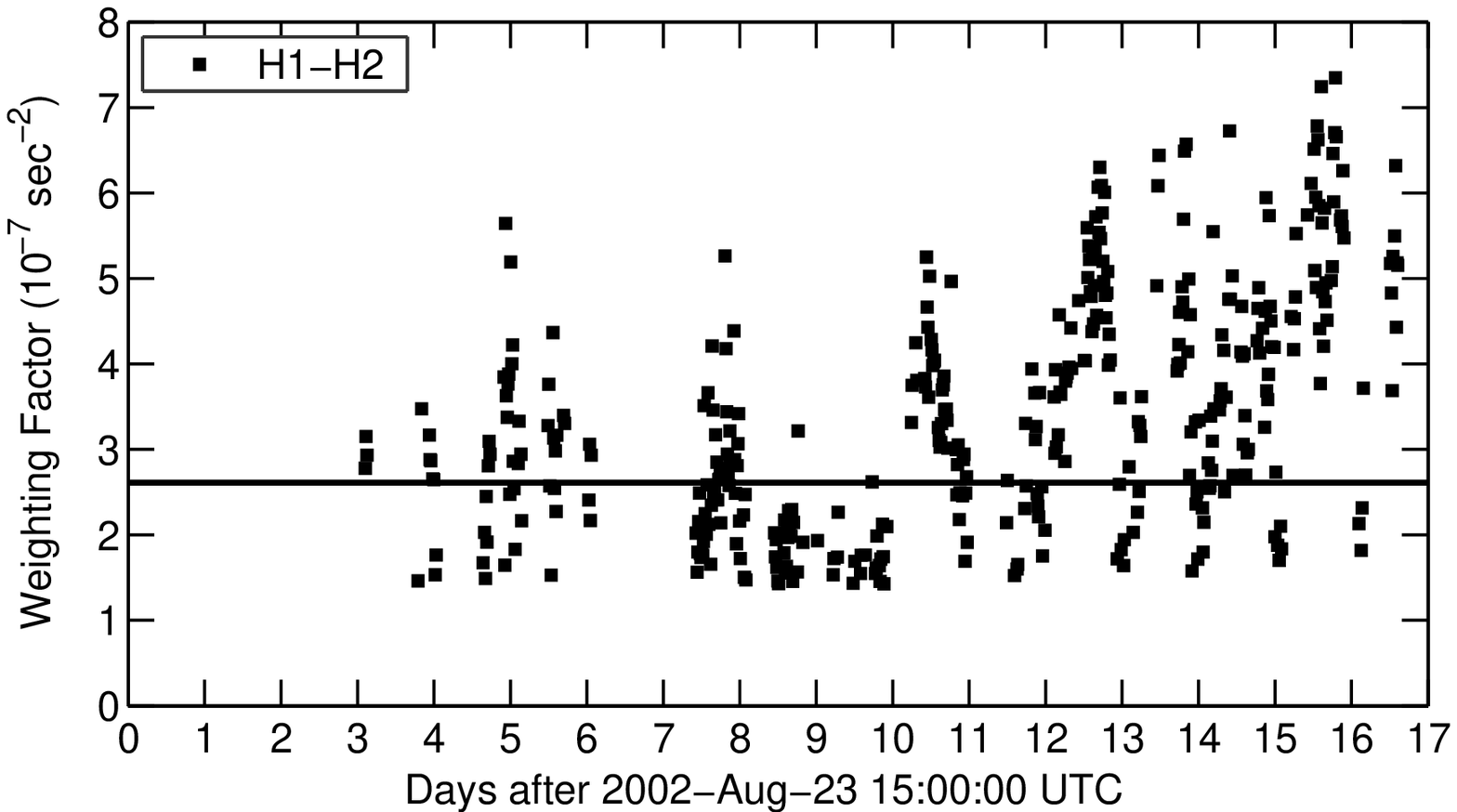}
  \caption{The weighting factors $\sigma_{Y_{IJ}}^{-2}$ for each 
  interferometer
  pair over the course of the S1 run; each point represents 900-sec of data.
  In each plot, a horizontal line indicates the weighting factor corresponding
  to the detector power spectra averaged over the whole run.}
  \label{fig:s1weights}
\end{figure}

Shown in Fig.~\ref{fig:s1weights} are the weighting factors
$\sigma_{Y_{IJ}}^{-2}$ (cf.\ Eq.~\ref{e:sigma2_discrete_final}) over 
the duration of the S1 run.
The $\sigma_{Y_{IJ}}^{-2}$ enter the expression for $Y_{\rm opt}$
(Eq.~\ref{e:Yoptimal}) and give a quantitative measure of the
sensitivity of a pair of detectors to a stochastic gravitational wave
background during the $I$-th interval.  Additionally, to gauge the
accuracy of the weighting factors, we compared the theoretical
standard deviations $\sigma_{Y_{IJ}}$ to the measured standard deviations
$s_{Y_{IJ}}$ (c.f.\ Eq.~\ref{e:s_I}).  For each interferometer pair, all 
but one or two of the $\sigma_{Y_{IJ}}/s_{Y_{IJ}}$ ratios lie between 
$0.5$ and 2, and show no systematic trend above or below unity.

\begin{figure}[htbp!]
  \includegraphics[width=3.5in,angle=0]{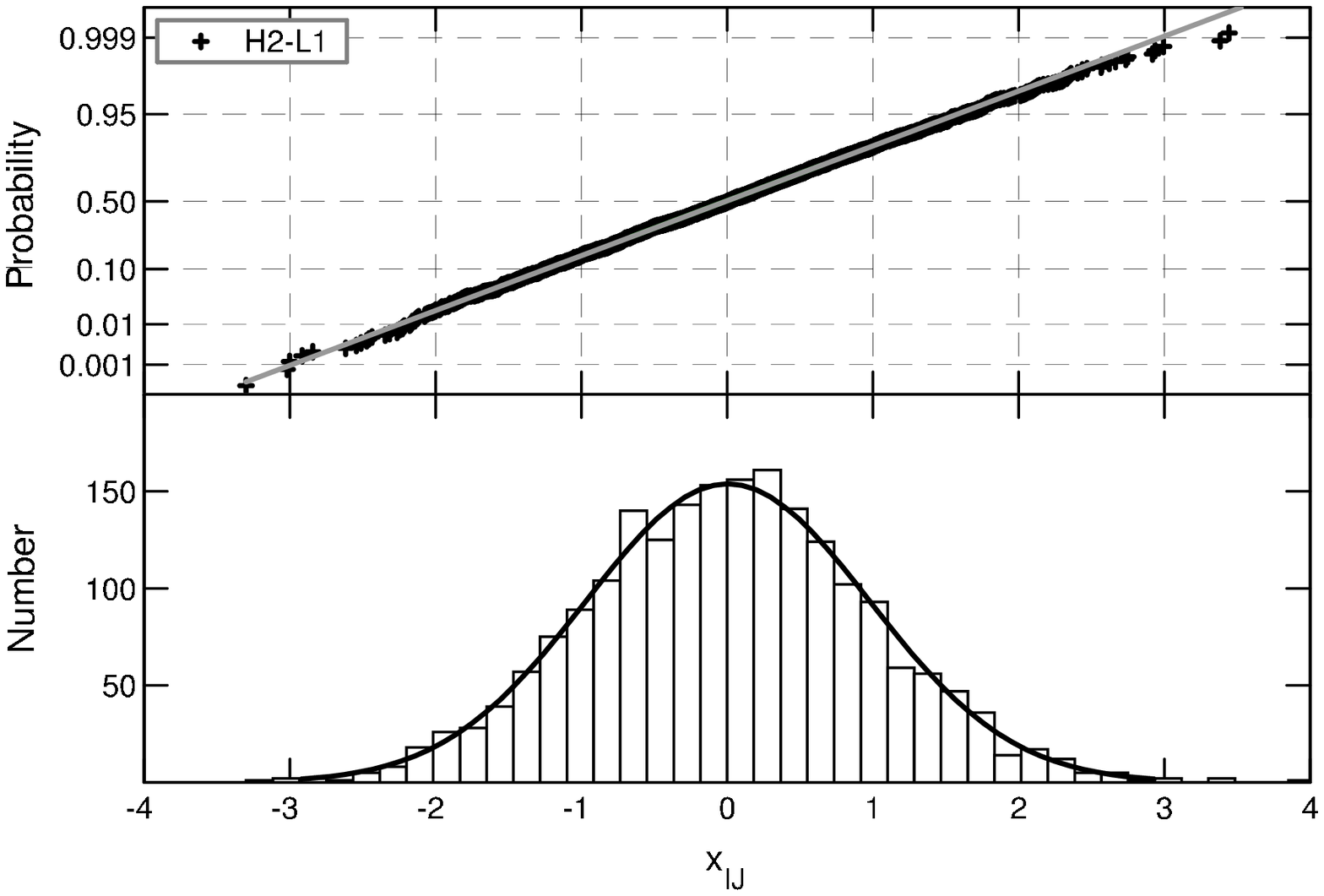}
  \includegraphics[width=3.5in,angle=0]{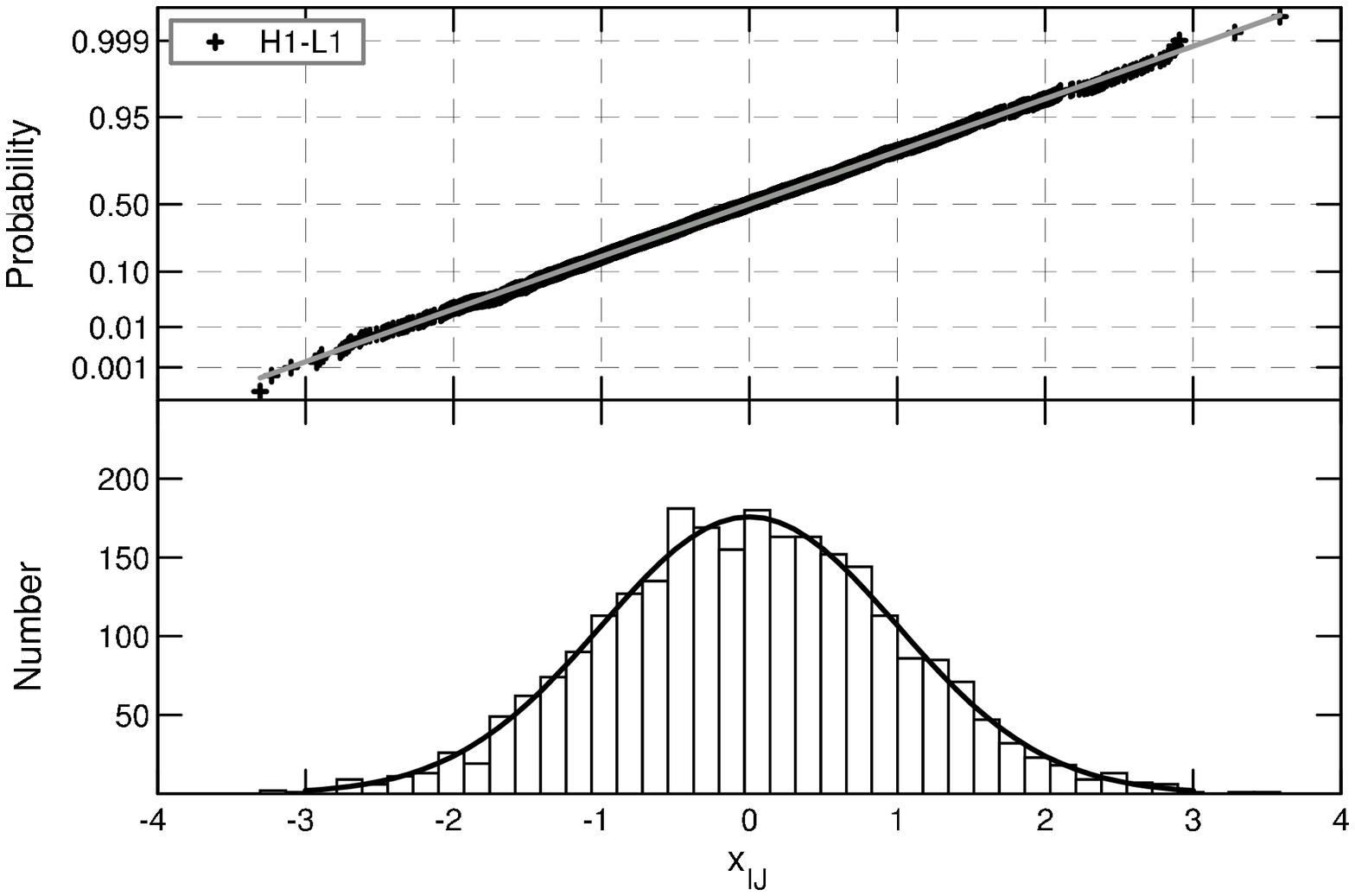}
  \includegraphics[width=3.5in,angle=0]{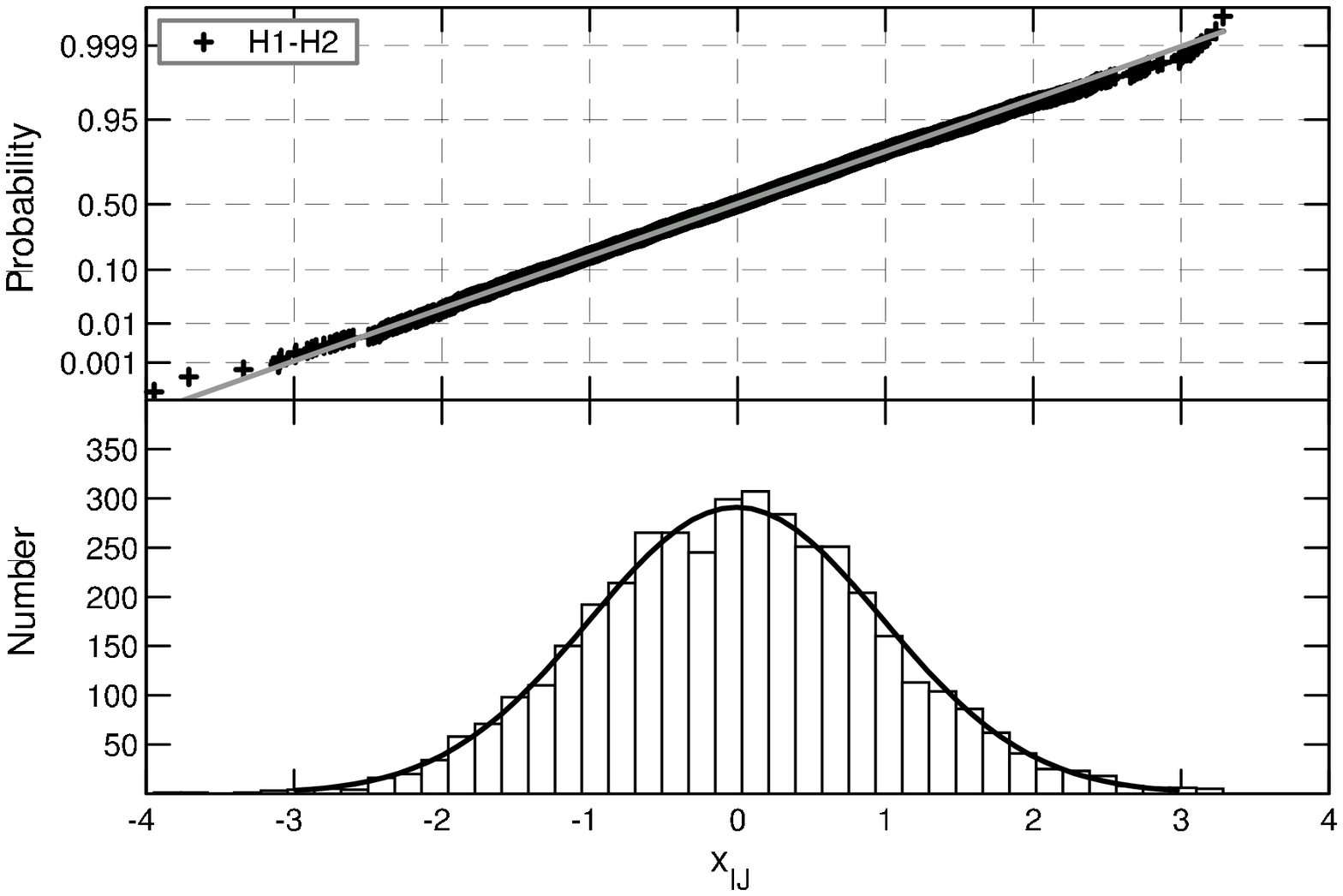}
  \caption{Normal probabilities and histograms of the values
    $x_{IJ}\equiv(Y_{IJ}-Y_{\rm opt})/\sigma_{Y_{IJ}}$, for all ${I,J}$ 
    included
    in the Table~\ref{tab:production_results} results. In theory,
    these values should be drawn from a Gaussian distribution of
    zero mean and unit variance.  The solid lines indicate the
    Gaussian that best fits the data; in the cumulative probability plots,
    curvature away from the straight lines is a sign of non-Gaussian
    statistics.}
  \label{fig:probabilityplots}
\end{figure}

Finally, Fig.~\ref{fig:probabilityplots} shows the distribution of the
cross-correlation values with mean removed and normalized by the
theoretical standard deviations---i.e.,
$x_{IJ}\equiv(Y_{IJ}-Y_{\rm opt})/\sigma_{Y_{IJ}}$. 
The values follow quite closely the expected Gaussian distributions.
%

\subsection{Frequency- and time-domain characterization}
\label{sec:fdomain}

\begin{figure}[htbp!]
  \includegraphics[width=3.5in,angle=0]{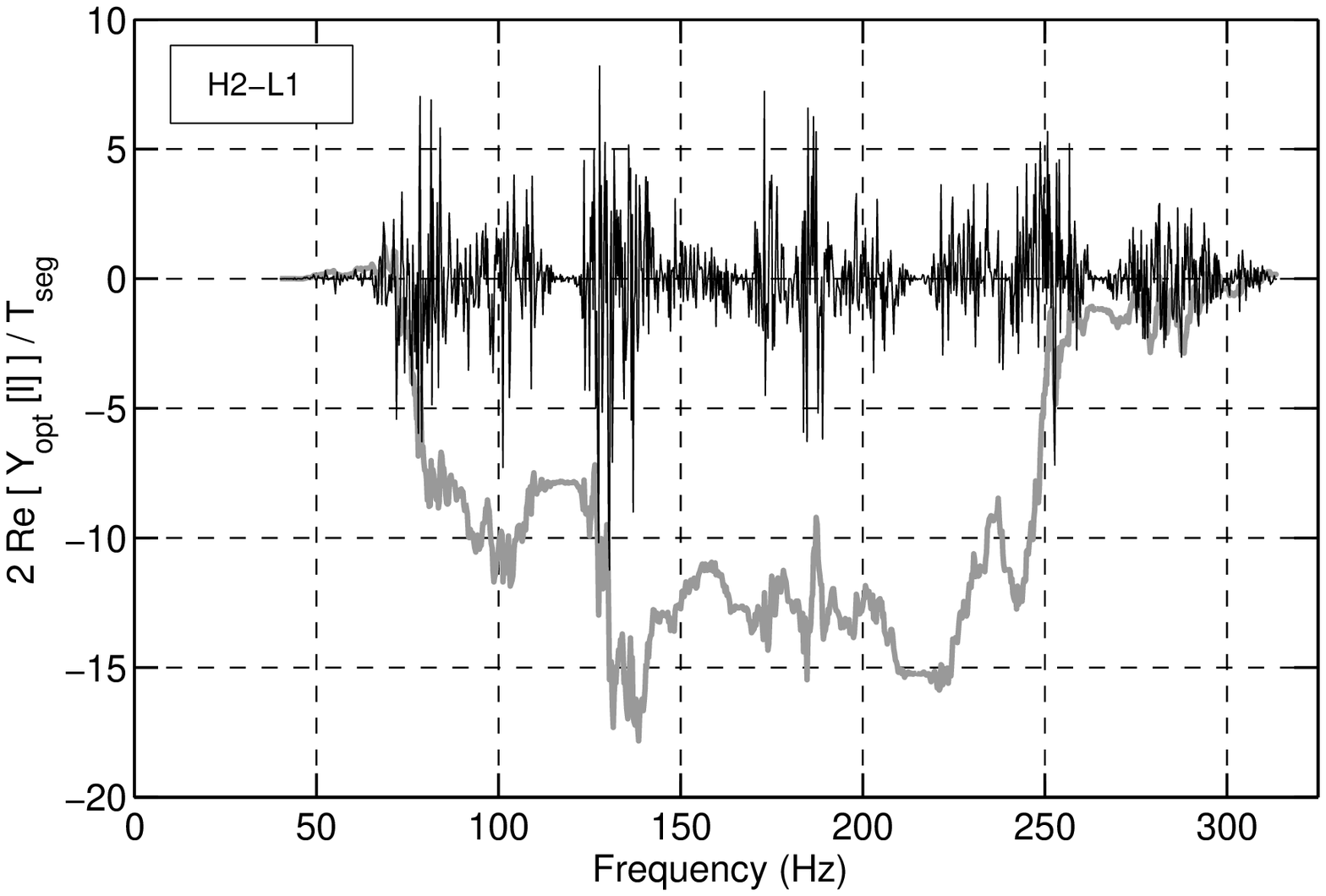}
  \includegraphics[width=3.5in,angle=0]{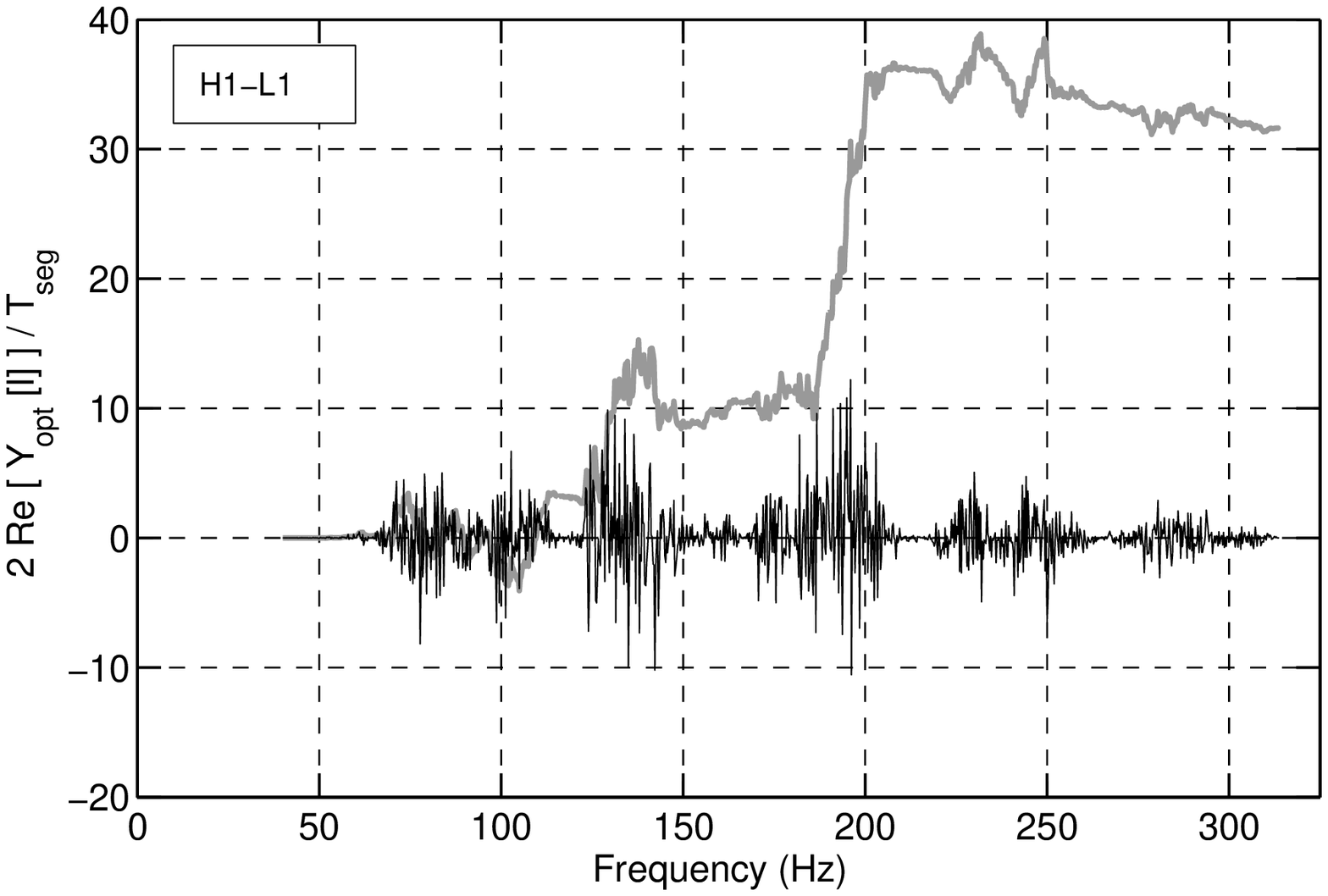}
  \includegraphics[width=3.5in,angle=0]{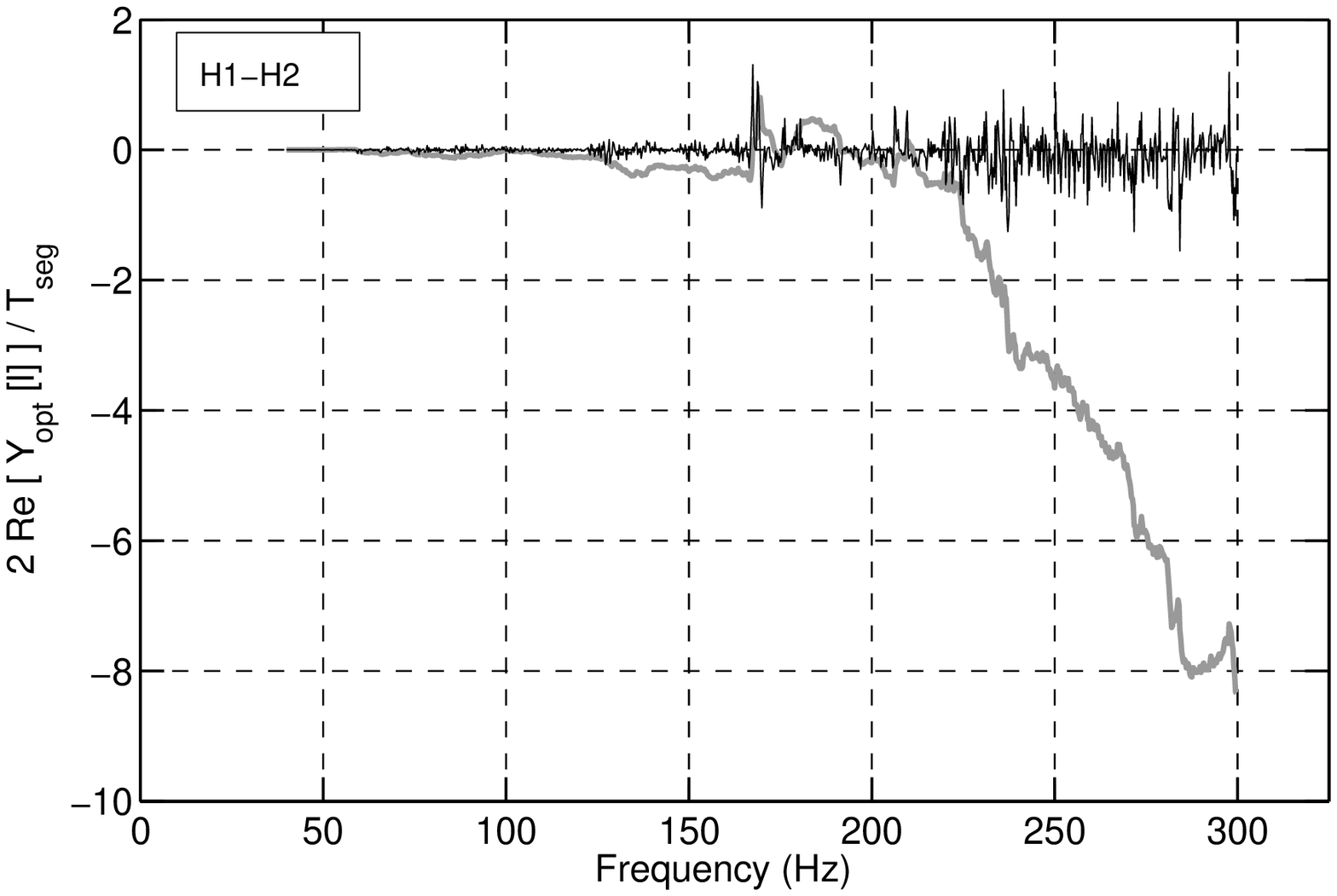}
  \caption{Real part of the cross-correlation spectrum,
  $\widetilde Y_{\rm opt}[\ell]/T_{\rm seg}$ (units of Hz$^{-1}$),
  for each detector pair.  The grey line in each plot shows the
  cumulative sum of the spectrum from 40~Hz to $f_\ell$, multiplied
  by $\delta f$ (which makes it dimensionless); the value of this
  curve at the right end is our estimate
  $\widehat\Omega_{\rm eff}\,h_{100}^2$.
  Note that the excursions in the cumulative sum for the H1-L1
  and H2-L1 correlations have magnitudes less than 1-2 error 
  bars of the corresponding point estimates; simulations 
  with uncorrelated data show the same qualitative behavior.}
  \label{fig:ccspectra}
\end{figure}

Because of the broadband nature of the interferometer strain data,
it is possible to explore the frequency-domain character of the
cross-correlations. In the analysis pipeline, we keep track of the
individual frequency bins that contribute to each $Y_{IJ}$, and
form the weighted sum of frequency bins over the full processed
data to produce an aggregate cross-correlation spectrum,
$\widetilde Y_{\rm opt}[l]$, for each detector pair 
(c.f.\ Eq.~\ref{e:Y_opt[l]}).
These spectra, along with their cumulative sums, are shown in
Fig.~\ref{fig:ccspectra}.
$\widetilde Y_{\rm opt}[\ell]$ can be quantitatively compared to the 
theoretically expected signal arising from a stochastic background
with $\Omega_{\rm gw}(f)\equiv\Omega_0={\rm const}$ by forming 
the $\chi^2$ statistic:
\begin{equation}
\chi^2(\Omega_0)\equiv\sum_{\ell=\ell_{\rm min}}^{\ell_{\rm max}} 
\frac{\left[{\rm Re}(\,\widetilde Y_{\rm opt}[\ell]\,) -
\mu_{\widetilde Y_{\rm opt}}[\ell]\right]^2}
{\sigma_{\widetilde Y_{\rm opt}}^2[\ell]}\,,
\label{e:chisquare}
\end{equation}
which is a quadratic function of $\Omega_0$. The sum runs over the 
$\sim\!1000$ frequency bins%
\footnote{To be exact, 1020 frequency bins were used for the
H1-L1, H2-L1 correlations and 1075 bins for H1-H2.} 
contained in each spectra. The expected 
values $\mu_{\widetilde Y_{\rm opt}}[\ell]$ and theoretical variances 
$\sigma_{\widetilde Y_{\rm opt}}^2[\ell]$ are given by
\begin{eqnarray}
\mu_{\widetilde Y_{\rm opt}}[\ell]&\equiv&
\Omega_0\,
\frac{T_{\rm seg}}{2}\,
\frac{3 H_0^2}{10\pi^2}\,
{\overline{w_1w_2}}\,\times
\nonumber\\
&\times&\,
\frac{\sum_I \sigma_{Y_{IJ}}^{-2}\, 
{\cal N}_I\,
\frac{\gamma^2[\ell]}{f_\ell^6 P_{1I}[\ell] P_{2I}[\ell]}}
{\sum_I \sigma_{Y_{IJ}}^{-2}}\,, 
\label{e:mu_Yopt}\\
\sigma_{\widetilde Y_{\rm opt}}^2[\ell]&\equiv&
\frac{1}{10}\,
\frac{T_{\rm seg}}{4}\,
{\overline{w_1^2w_2^2}}\,\times
\nonumber\\
&\times&\
\frac{\sum_I \sigma_{Y_{IJ}}^{-4}\,
{\cal N}_I^2\,
\frac{\gamma^2[\ell]}{f_\ell^6 P_{1I}[\ell] P_{2I}[\ell]}}
{\left(\sum_I \sigma_{Y_{IJ}}^{-2}\right)^2}\,.
\label{e:sigma2_Yopt}
\end{eqnarray}
Note that by
using Eqs.~\ref{e:normalization_discrete},
\ref{e:sigma2_discrete_final}, one can show
\begin{eqnarray}
&&2\sum_{\ell=\ell_{\rm min}}^{\ell_{\rm max}} 
\delta\!f\ \mu_{\widetilde Y_{\rm opt}}[\ell] = 
\Omega_0\,h_{100}^2\, T_{\rm seg}\,,
\\
&&2\sum_{\ell=\ell_{\rm min}}^{\ell_{\rm max}} 
\delta\!f\ \sigma^2_{\widetilde Y_{\rm opt}}[\ell] =
\frac{1}{10}\,
\left(\sum_I\sigma_{Y_{IJ}}^{-2}\right)^{-1}\,,
\end{eqnarray}
which are the expected value 
and theoretical variance 
of the weighted cross-correlation $Y_{\rm opt}$ 
(c.f.\ Eq.~\ref{e:Yoptimal}).

For each detector pair, we find that the minimum $\chi^2$ value 
occurs at the corresponding estimate 
$\widehat\Omega_{\rm eff}\,h_{100}^2$ for that
pair, and the width of the $\chi^2 = \chi_{\rm min}^2 \pm 2.71$ 
points corresponds to the 90\% confidence intervals given in
Table~\ref{tab:production_results}.  For the H1-L1 and H2-L1 pairs,
the minimum values are $\chi_{\rm min}^2 =(0.96, 1.0)$ per
degree-of-freedom. This results from the low signal-to-noise ratio of
the measurements: $\widehat\Omega_{\rm eff}\,h_{100}^2/
\widehat\sigma_{\Omega,{\rm tot}} = (1.8,.0094)$.

For the H1-H2 pair, $\chi_{\rm min}^2=4.9$ per degree-of-freedom. In
this case the magnitude of the cross-correlation signal-to-noise ratio 
is relatively high, $\widehat\Omega_{\rm eff}\,h_{100}^2/
\widehat\sigma_{\Omega,{\rm tot}} = -8.8$, and the value of $\chi_{\rm
min}^2$ indicates the very low likelihood that the measurement is
consistent with the stochastic background model.  For $\sim\!1000$
frequency bins (the number of degrees-of-freedom of the fit), the
probability of obtaining such a high value of $\chi_{\rm min}^2$ is
extremely small, indicating that the source of the observed H1-H2
correlation is not consistent with a stochastic gravitational wave
background having $\Omega_{\rm gw}(f)\equiv\Omega_0={\rm const}$.

\begin{figure*}[htbp!]
  \includegraphics[width=7.0in,angle=0]{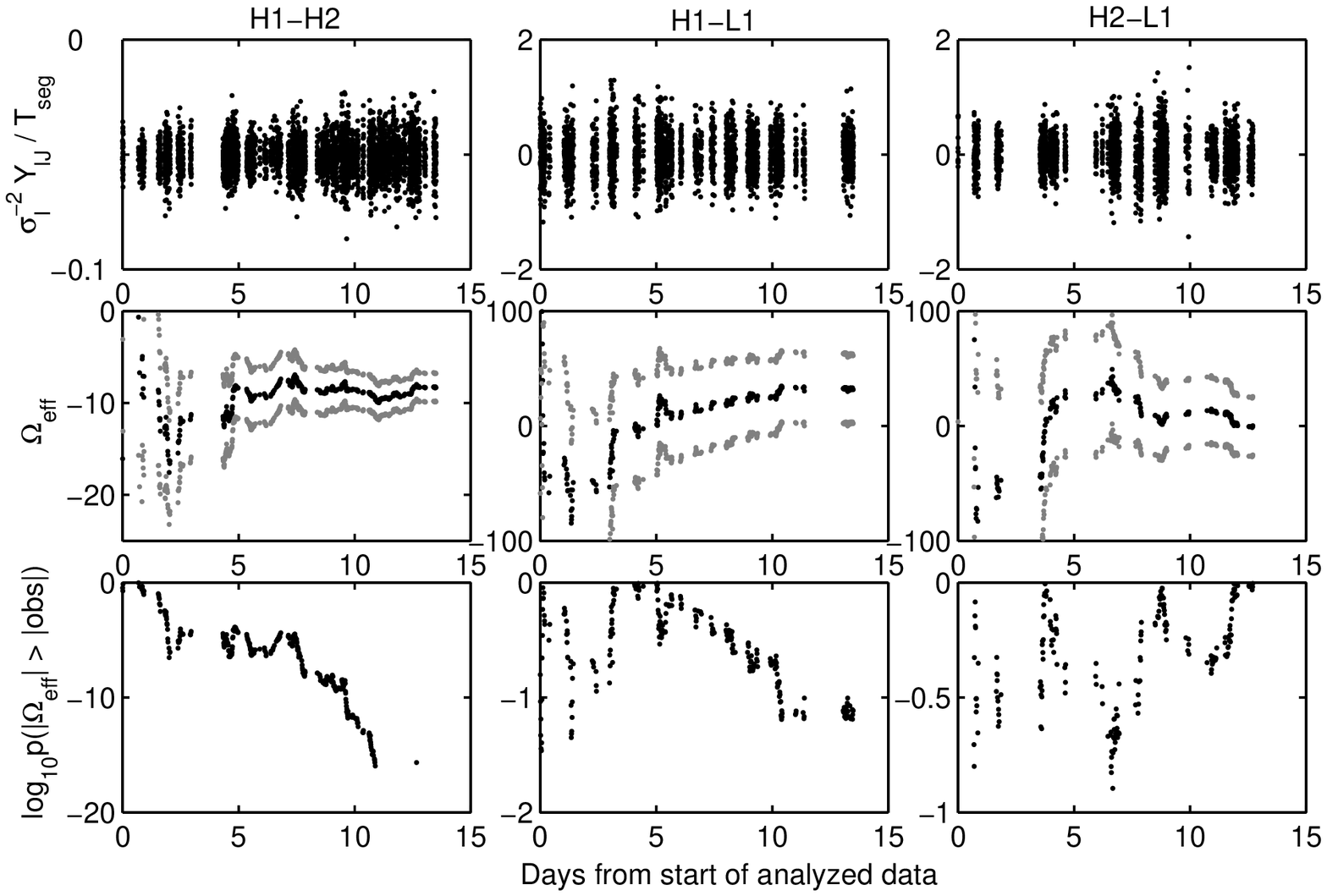}
  \caption{Cross-correlation statistics as a function of amount of
  data analyzed. Each column of plots shows the analysis results for
  a given detector pair, as indicated, over the duration of the data
  set. {\bf{Top plots:}} Points correspond to the cross-correlation
  statistic values $Y_{IJ}$ appropriately normalized,
  $M\,T_{\rm seg}^{-1}\,\sigma_{Y_{IJ}}^{-2}\,Y_{IJ}/
    \sum_I \sigma_{Y_{IJ}}^{-2}$,
  where $M$ is the total number of analyzed intervals,
  so that the mean of all the values is the final point estimate
  $\widehat\Omega_{\rm eff}\,h_{100}^2$.
  The scatter shows the variation in the point
  estimates from segment to segment. {\bf{Middle plots:}} Evolution
  over time of the estimated value of $\Omega_{\rm eff}\,h_{100}^2$.
  The black points give the estimates 
  $\widehat\Omega_{\rm eff}\,h_{100}^2$
  and the grey points give the
  $\pm1.65\,\widehat\sigma_\Omega$ errors ($90\%$ confidence bounds),
  where $\widehat\sigma_\Omega$ is defined by Eq.~\ref{e:sigma_Omega}.
  The errors decrease with time, as expected from a $T^{-1/2}$
  dependence on observation time. {\bf{Bottom plots:}} Assuming that
  the estimates shown in the middle plots are drawn from zero-mean Gaussian
  random variables with the error bars indicated, the probability of
  obtaining a value of $|\Omega_{\rm eff}\,h_{100}^2| \ge$ observed
  absolute value is given by:
  $p(|\Omega_{\rm eff}\,h_{100}^2|\ge |x|) = 1 -
  \textrm{erf}({|x|/\sqrt{2}\,\widehat\sigma_\Omega})$.
  This is
  plotted in the bottom plots. For the H1-H2 pair, the probability
  becomes $< 10^{-20}$ after approximately 11 days.
  While the H1-L1 pair shows a
  signal-to-noise ratio above unity,
  $\widehat\Omega_{\rm eff}\,h_{100}^2/ 
  \widehat\sigma_{\Omega,{\rm tot}} = 1.8$,
  there is a 10\% probability of
  obtaining an equal or larger value from random noise alone.}
  \label{fig:statsvstime}
\end{figure*}

It is also interesting to examine how the cross-correlation 
behaves as a function of the volume of data analyzed.
Figure~\ref{fig:statsvstime} shows the weighted cross-correlation
statistic values versus time, and the evolution of the estimate 
$\widehat\Omega_{\rm eff}\,h_{100}^2$ and statistical error bar
$\widehat\sigma_\Omega$ over the data run.
Also plotted are the probabilities $p(|\Omega_{\rm eff}\,h_{100}^2|\ge
|x|) = 1 - \textrm{erf}({|x|/\sqrt{2}\,\widehat\sigma_\Omega})$ of 
obtaining a value of $\Omega_{\rm eff}\,h_{100}^2$ greater than or 
equal to the observed value, assuming that these values are drawn from 
a zero-mean Gaussian random distribution, of width equal to the 
cumulative statistical error at each point in time.  For the H1-L1 and
H2-L1 detector pairs, the probabilities are $\agt 10\%$ for the
majority of the run.  For H1-H2, the probability drops below
$10^{-20}$ after about 11 days, suggesting the presence of a non-zero
instrumental correlation (see also Sec.~\ref{sec:H1H2correlation}).
%

\subsection{Time shift analysis}

\begin{figure}[htbp!]
  \includegraphics[width=3.5in,angle=0]{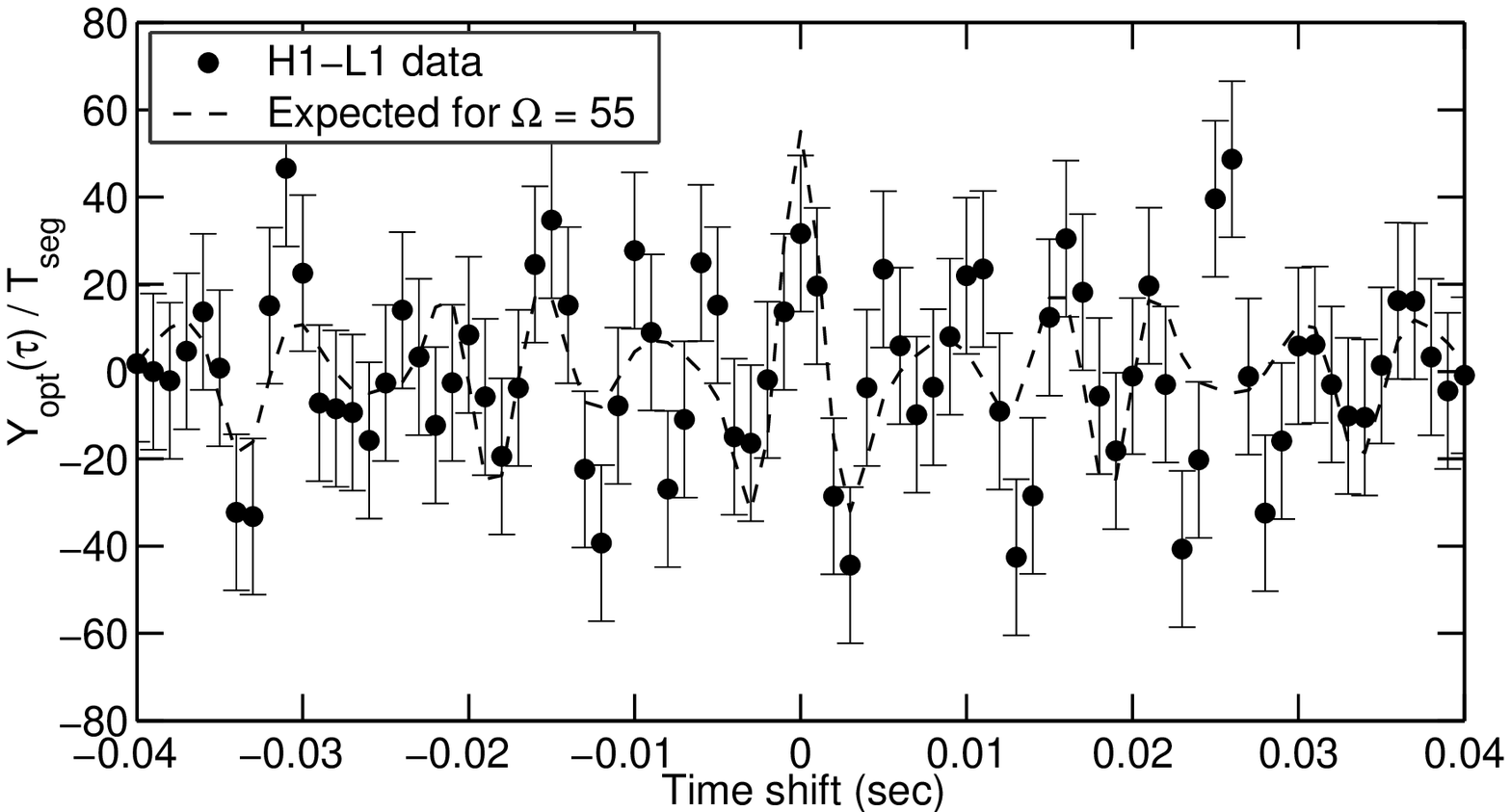}
  \includegraphics[width=3.5in,angle=0]{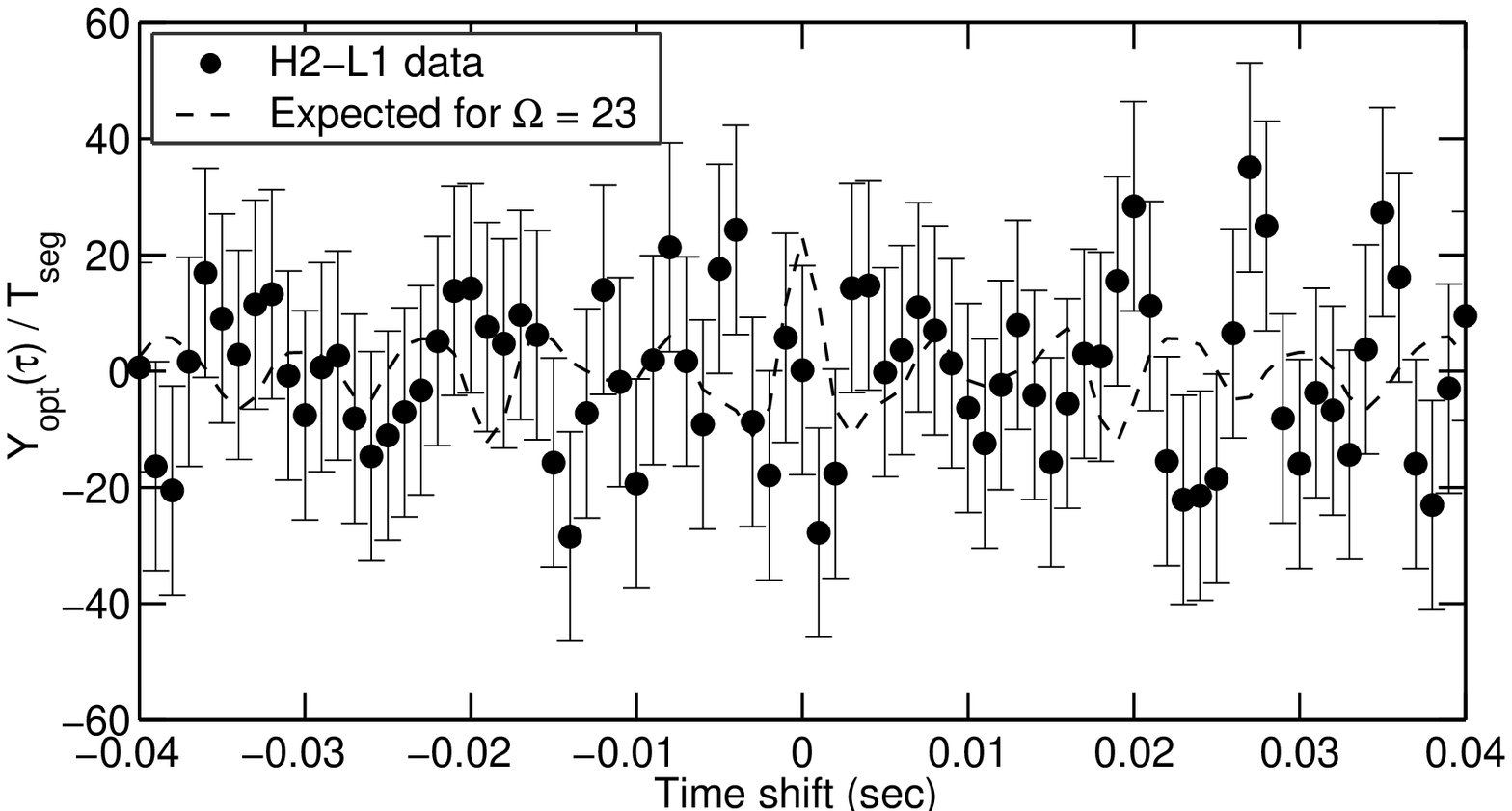}
  \includegraphics[width=3.5in,angle=0]{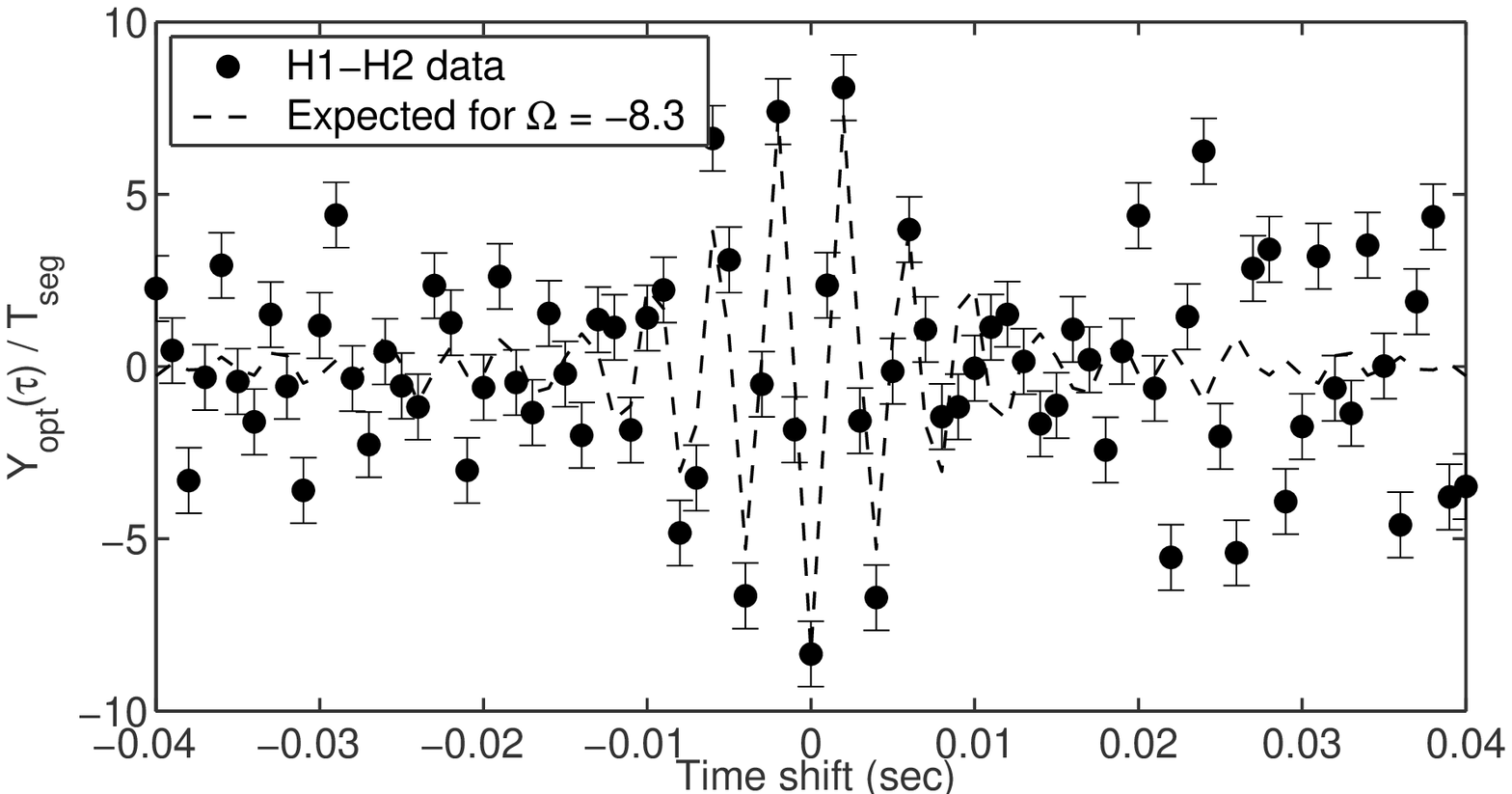}
  \caption{Results of a time shift analysis for the three detector pairs.
  Plotted are the discrete inverse Fourier transforms of 
  $\widetilde Y_{\rm opt}[l]$.  
  Also shown are the expected time shift curves in the presence of 
  a stochastic background with 
  $\Omega_{\rm gw}(f)\equiv\Omega_0={\rm const}$; 
  for the H1-L1 and H2-L1 pairs, these are computed using the
  corresponding upper limit levels $\Omega_0\,h_{100}^2 = 55$ and
  $\Omega_0\,h_{100}^2 = 23$, respectively, while for H1-H2 the instrumental
  correlation level of $-8.3$ is used.}
  \label{fig:timeshiftplots}
\end{figure}

It is instructive to examine the behavior of the
cross-correlation as a function of a relative time
shift $\tau$ introduced between the two data streams:
%
\begin{eqnarray}
Y(\tau) &\equiv& \int_{-T/2}^{T/2} dt_1\int_{-T/2}^{T/2} dt_2\,
s_1(t_1 - \tau)\,Q(t_1-t_2)\, s_2(t_2) \nonumber\\
&=& \int_{-\infty}^{\infty}df\ e^{i 2 \pi f \tau}\,\widetilde Y(f)\,,
\label{e:YofTau}
\end{eqnarray}
where
\begin{equation}
\widetilde Y(f)\equiv \int_{-\infty}^{\infty}df'\
\delta_T(f-f')\,\tilde s_1^*(f)\,
\tilde Q(f')\,\tilde s_2(f')\,.
\label{e:Yoff}
\end{equation}
Thus, $Y(\tau)$ is simply the inverse Fourier transform of the
integrand, $\widetilde Y(f)$, of the cross-correlation statistic $Y$
(c.f.\ Eq.~\ref{e:Y_continuous_freq}).  The discrete frequency version
of this quantity, $\widetilde Y_{\rm opt}[\ell]$ (c.f.\
Eq.~\ref{e:Y_opt[l]}), is shown in Fig.~\ref{fig:ccspectra} for each
detector pair.  Figure~\ref{fig:timeshiftplots} shows the result of
performing discrete inverse Fourier transforms on these spectra.  For
time shifts very small compared to the original FFT data length of
90-sec, this is equivalent to shifting the data and recalculating the
point estimates. 

Also shown are expected time shift curves in the
presence of a significant stochastic background with 
$\Omega_{\rm gw}(f)\equiv\Omega_0={\rm const}$.  These are obtained 
by taking the inverse discrete
Fourier transforms of Eq.~\ref{e:mu_Yopt}; they have an oscillating
behavior reminiscent of a sinc function.  For the LHO-LLO pairs, these
are computed for the upper limit levels given in
Table~\ref{tab:production_results}, while for H1-H2, the expected
curve is computed taking $\Omega_0\,h_{100}^2$ equal to the 
instrumental correlation level of $-8.3$.  For the two inter-site
correlations (H1-L1 and H2-L1), most of the points lie within the
respective standard error levels: $\widehat\sigma_{\Omega,{\rm tot}} =
18$ for H1-L1, and $\widehat\sigma_{\Omega, {\rm tot}} = 18$ for
H2-L1; for H1-H2, most of the points lie outside the error level,
$\widehat\sigma_{\Omega, {\rm tot}} = 0.95$, indicating once again
that the observed H1-H2 correlation is inconsistent with the 
presence of a stochastic background with
$\Omega_{\rm gw}(f)\equiv\Omega_0={\rm const}$.

\section{ERROR ESTIMATION}
\label{sec:errors}

We have identified three potentially significant types of error that
contribute to the total error on our estimate of
$\Omega_{\rm eff}\,h_{100}^2$.
The first is a theoretical statistical error,%
\footnote{Here we are treating $\widehat\Omega_{\rm eff}\,h_{100}^2$,
$Y_{\rm opt}$, and $Y_I$ as {\em random variables} and not as
their values for a particular realization of the data.}
\begin{equation}
\sigma_\Omega\equiv
\sigma_{Y_{\rm opt}/T_{\rm seg}}
=\frac{1}{T_{\rm seg}}
\frac{\sqrt{\sum_I\sigma_{Y_{IJ}}^{-4}\sigma_{Y_I}^2}}
{\sum_I\sigma_{Y_{IJ}}^{-2}}
\label{e:sigma_Omega}
\end{equation}
where the second equality follows from the definition of
$\widehat\Omega_{\rm eff}\,h_{100}^2$
in terms of $Y_{\rm opt}$ and $Y_I$, treating
the weighting factors $\sigma^{-2}_{Y_{IJ}}$ as constants in the
calculation of the theoretical variance of $Y_{\rm opt}$.
We estimate this error by replacing the theoretical variance
$\sigma_{Y_I}^2 (=\sigma^2_{Y_{IJ}}/10)$ by its unbiased estimator
$s^2_{Y_{IJ}}/10$ (c.f.\ Eqs.~\ref{e:Y_I}, \ref{e:s_I}).%
\footnote{This is valid provided the individual cross-correlation
measurements $Y_{IJ}$ are statistically independent of one
another.  This assumption was tested by computing the auto-covariance
function of the $Y_{IJ}$ data sequences; for each of the three
$Y_{IJ}$ sets, the result was a delta function at zero-lag, as
expected for independent data samples.}
Thus,
\begin{equation}
\widehat\sigma_\Omega\equiv\frac{1}{T_{\rm seg}}\frac{1}{\sqrt{10}}
\frac{\sqrt{\sum_I\sigma_{Y_{IJ}}^{-4} s_{Y_{IJ}}^2}}
{\sum_I\sigma_{Y_{IJ}}^{-2}}\,.
\label{e:hat_sigma_Omega}
\end{equation}

The last two sources of error are due to unresolved time variations 
in the interferometers' calibration, $\sigma_{\Omega,{\rm cal}}$, 
and strain noise power spectra, $\sigma_{\Omega,{\rm psd}}$. 
As described earlier,
detector power spectrum estimates are made on 900-sec data intervals,
and a single response function, derived from the central 60-seconds of
calibration line data, is applied to each interval. Variations in both
the response functions and the power spectra occur on shorter time
scales, and we have estimated the systematic errors
($\widehat\sigma_{\Omega,{\rm cal}}$ and $\widehat\sigma_{\Omega,{\rm
psd}}$) due to these variations as follows. The cross-correlation
analysis is performed again using a finer time resolution for
calibration and power spectrum estimation, and the results are assumed
to be representative of the effect of variations at other time
scales. Specifically, each detector pair is re-analyzed with power
spectrum estimates, and corresponding optimal filters, computed for
each 90-sec data segment (using the same frequency resolution as the
original analysis, but approximately $1/10$ the number of
averages). Separately, each detector pair is also re-analyzed using
the calibration line amplitudes, and resulting response functions,
corresponding to each 90-sec data segment. Each analysis yields a new
point estimate $\widehat\Omega_{\rm eff}\,h_{100}^2$; 
for each re-analysis, the
difference between the new point estimate and the original point
estimate is used as the estimate of the systematic errors
$\widehat\sigma_{\Omega,{\rm cal}}$ and $\widehat\sigma_{\Omega,{\rm
psd}}$.  The total error is then formed as:
\begin{equation}
\widehat\sigma^2_{\Omega,{\rm tot}} = 
\widehat\sigma^2_{\Omega} + \widehat\sigma^2_{\Omega,{\rm cal}} +
\widehat\sigma^2_{\Omega,{\rm psd}}.
\label{e:hat_sigma2_Omega_tot}
\end{equation}

These error estimates are shown in
Table~\ref{tab:errorestimates}. Also shown in the table are
values for the fractional calibration uncertainty. The
significant effect here is a frequency-independent uncertainty in
the response function magnitude; uncertainties in the phase
response are negligible in the analysis band.

\begin{table*}
\begin{ruledtabular}
\begin{tabular}{cccccc}
&&&&& Calibration\\
Pair & $\widehat\sigma_{\Omega}$ & $\widehat\sigma_{\Omega,{\rm psd}}$ & 
$\widehat\sigma_{\Omega,{\rm cal}}$ &
$\widehat\sigma_{\Omega,{\rm tot}}$ & uncertainty\\
\hline
H1-H2 & 0.93 & 0.078 & 0.16 & 0.95 & $\pm 20\%$\\
H1-L1 & 18 & 0.23 & 0.29 & 18 & $\pm 20\%$\\
H2-L1 & 15 & 9.3 & 1.2 & 18 & $\pm 20\%$\\
\end{tabular}
\end{ruledtabular}
\caption{Sources of error in the estimate $\widehat\Omega_{\rm
eff}\,h_{100}^2 = Y_{\rm opt}/T_{\rm seg}$: 
$\widehat\sigma_{\Omega}$ is the statistical error; 
$\widehat\sigma_{\Omega,{\rm psd}}$ is the
error due to unresolved time variations of the equivalent strain
noise in the detectors; and $\widehat\sigma_{\Omega,{\rm cal}}$ is 
the error due to unresolved calibration variations. The calibration
uncertainty for each detector pair results from adding linearly a
$\pm 10\%$ uncertainty for each detector, to allow for a worst
case combination of systematic errors.} \label{tab:errorestimates}
\end{table*}

We have also considered the effect of data acquisition system
timing errors on the
analysis. The behavior of the cross-correlation statistic when a
time offset is introduced into the analysis was shown in
Fig.~\ref{fig:timeshiftplots}. A finer resolution plot of the
time-shift curve in the presence of a significant stochastic
background indicates that a $\tau = \pm 400~\mu \rm{sec}$
offset between the LHO and LLO interferometers corresponds to a
$10\%$ reduction in the estimate of our upper limit. The growth of
this error is roughly quadratic in $\tau$. Throughout the S1
run, the time-stamping of each interferometer's data was
monitored, relative to GPS time. The relative timing error between
H2 and L1 was approximately $40~\mu{\rm sec}$ for roughly half the
analyzed data set, $320~\mu{\rm sec}$ for 32\% of the data set,
and $600~\mu{\rm sec}$ for 16\% of the set. The combined effect of
these timing offsets is an effective reduction in the point
estimate, $\widehat\Omega_{\rm eff}\,h_{100}^2$, 
of 3.5\%, a negligible effect.
The H1-L1 relative timing errors were even smaller, being less
than $30~\mu{\rm sec}$ during the whole data set.

\section{VALIDATION: SIGNAL INJECTIONS}
The analysis pipeline was validated by demonstrating the ability
to detect coherent excitation of the interferometer pairs
produced by simulated signals corresponding to a stationary,
isotropic stochastic gravitational wave background.

A software package was developed to generate a pseudorandom time
series representing this excitation. In this manner, pairs of
coherent data trains of simulated stochastic signals could be
generated. The amplitude of the simulated stochastic background
signal was adjusted by an overall scale factor, and the behavior
of the detection algorithm could be studied as a function of
signal-to-noise ratio. Simulated data were either injected into
the interferometer servo control system in order to directly
stimulate the motion of the interferometer mirrors ({\em{hardware
injections}}) or the calculated waveforms could be added in
software to the interferometer data as part of the analysis
pipeline ({\em{software injections}}). The former approach was
used to inject a few simulated stochastic background signals of
different amplitudes during interferometer
calibrations at the beginning and end of the S1 run. The latter
approach was used after the S1 run during the data analysis
phase. Table\ \ref{tab:signal_injections} lists the different
injections that were used to validate our procedure.

\subsection{Hardware injected signals}
\label{sec:hardware}

Hardware injections required that the simulated data trains be
first convolved with the appropriate instrument response
functions. These pre-processed data trains were then injected
digitally into the respective interferometer servo control
systems.

Simulated stochastic background signals with $\Omega_{\rm
gw}(f)\equiv\Omega_0={\rm const}$ were injected simultaneously
into the H2-L1 pair for two 1024 second (17.07 min) periods
shortly after the S1 run was completed.  Referring to Table
\ref{tab:signal_injections}, the two injections had different
signal strengths, corresponding to signal-to-noise ratios of
$\sim20$ and $\sim10$, respectively. The stronger injection
produced a noticeable increase in the H2 power spectrum in the
band from 40 to 600~Hz.

\begin{table*}
\begin{ruledtabular}
\begin{tabular}{cccccc}

Interferometer & Hardware (HW) & Magnitude of Injected & Approx.
& Magnitude of Detected Signal \\
Pair & Software (SW) & Signal ($\Omega_0\,h_{100}^2$)& SNR
& (90{\% }CL, 
$\widehat\Omega_0\,h_{100}^2 \pm 1.65\,\widehat\sigma_\Omega$)\\
\hline
H2-L1 & HW & $3906$  & $10$ & $3744 \pm 663$   \\
H2-L1 & HW & $24414$ & $17$ & $25365 \pm 2341$ \\
\hline
H2-L1 & SW & $16$    & $-$  & $-$              \\
H2-L1 & SW & $100$   & $-$  & $-$              \\
H2-L1 & SW & $625$   & $3$  & $891 \pm 338$    \\
H2-L1 & SW & $3906$  & $13$ & $4361 \pm 514$   \\
H2-L1 & SW & $24414$ & $50$ & $25124 \pm 817$  \\

\end{tabular}
\end{ruledtabular}
\caption{Summary of injected signals used to validate the
analysis pipeline. Both hardware injections during the S1 run and
post-S1 software injections were used. Injections were introduced
into short data segments (refer to text). The signal-to-noise
ratios shown correspond to integration times that are much
shorter than the full S1 data set, and thus the lower
signal-to-noise ratio injections were not detectable. The
software and hardware injections have different signal-to-noise 
ratios for the same $\Omega_0\,h_{100}^2$ values
due to the variation in the interferometer noise power spectral 
densities at the different epochs when the signals were injected.}
\label{tab:signal_injections}
\end{table*}

In principle, the stochastic gravitational wave background
estimate can be derived from a single (point) measurement of the
cross-correlation between pairs of interferometers. However, in
order to verify that the simulated signals being detected were
consistent with the process being injected, time shift analyses of
the data streams were performed for a number of different offsets,
$\tau$.
This technique can potentially identify instrumental and
environmental correlations that are not astrophysical.

%
%

For each injection, the results of the time-shifted analysis were
compared with the expected time shift curves.
Allowing for possible (unknown) time shifts
associated with the stimulation and data acquisition processes, a
two parameter $\chi^{2}$ regression analysis was performed on the
time shift data to determine: (i) the time offset (if any existed),
(ii) the amplitude of the signal, and (iii) the uncertainties in
the estimation of these parameters.

\begin{figure}[htbp!]
  \includegraphics[width=3.5in,angle=0]{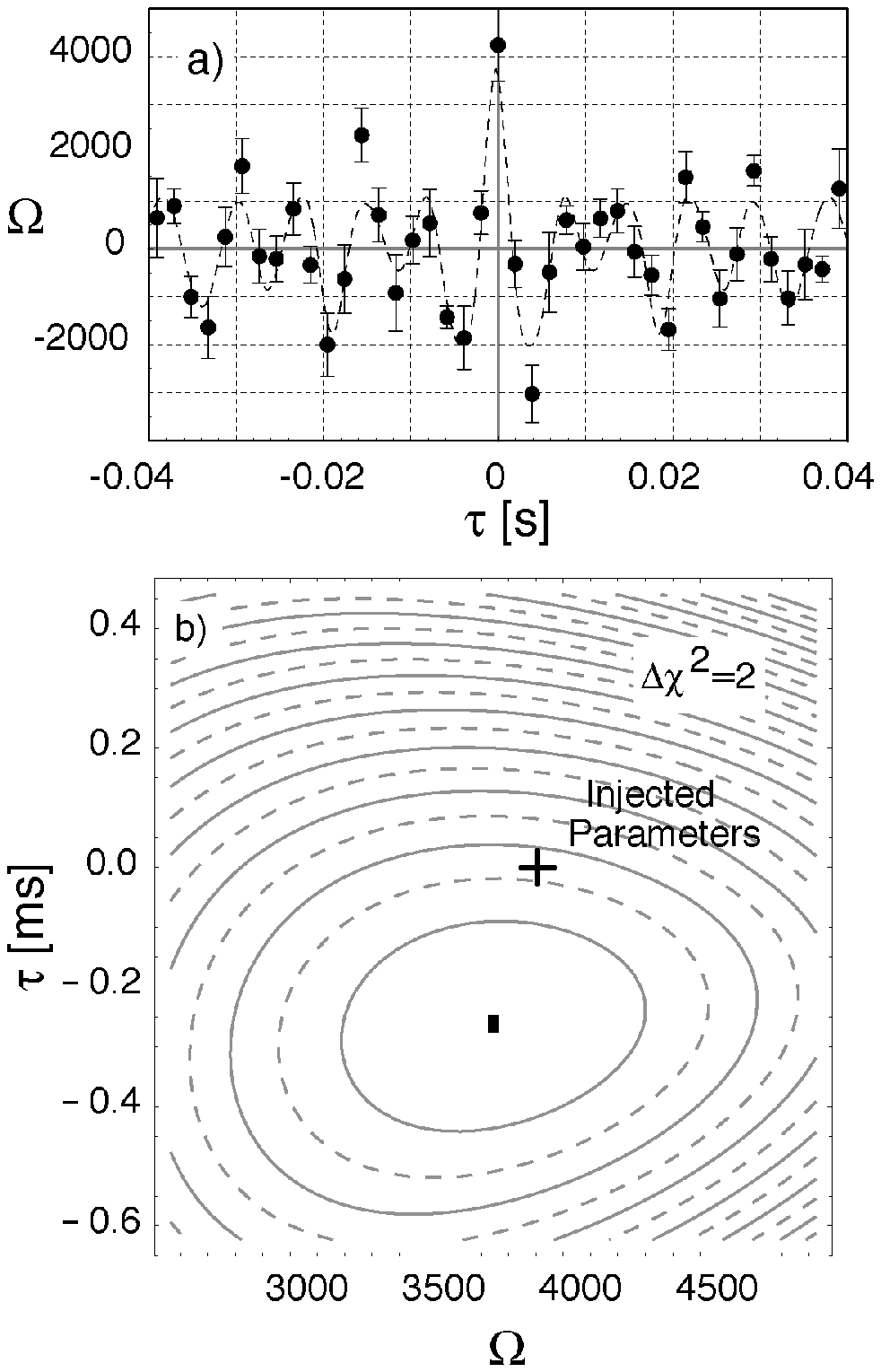}
  \caption{Hardware injection time shift analysis for the H2-L1
  interferometer pair, with signal strength $\Omega_0\,h_{100}^2=
  3906$. Panel (a): Time shift dependence of
    the cross-correlation (refer to Eq.~\ref{e:YofTau}).
    The data are shown with $\pm 1\widehat\sigma_\Omega$ error bars 
    estimated from the measured
    quantities (Eq.~\ref{e:sigma_Omega}) for each time offset.
    The dashed curve is the expected dependence, scaled and offset in
    time to provide a best fit. Panel (b): 
    Contour plot of $\chi^2(\Omega_0,\tau)$
    near the best fit. The minimum value is $\chi^2_{\rm min} = \rm{1.8}$
    (for 2 degrees of freedom), and occurs at the coordinates of
    the black rectangle: $\{\Omega_0\,h_{100}^2,\tau\} = 
    \{3744,\,-270~\mu{\rm sec} \}$.
    The cross (+) corresponds to the injected signal, whose estimated
    strength has 90\% confidence bounds of:
    $3345 \le \Omega_0\,h_{100}^2 \le 4142$. The best fit
    time offset of $-270~\mu$sec is within the observed relative data
    acquisition timing errors between H2 and L1 during S1.}
    \label{fig:chi2FitHW3906}
\end{figure}

Results of this analysis for the hardware injection with signal
strength $\Omega_0\,h_{100}^2=3906$ are shown in
Fig.~\ref{fig:chi2FitHW3906}.
The agreement between injected simulated signal and the
detected signal after end-to-end analysis with our pipeline
gave us confidence that the full data analysis pipeline was
working as expected.
%

\subsection{Software injected signals}
\label{sec:software}

\begin{figure}[htbp!]
  \includegraphics[width=3.5in,angle=0]{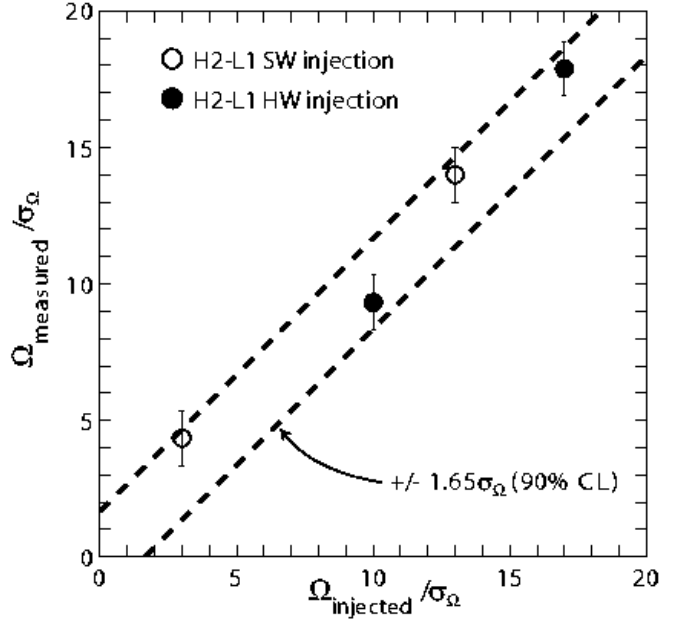}
  \caption{H2-L1 point estimates and error bars obtained from the S1 data
  analysis for both hardware and software injections. 
  Measured versus injected SNRs are shown for a number of simulations. 
  The ordinate of each point is the result of a $\chi^2$ analysis like 
  the one shown in Fig.~\ref{fig:chi2FitHW3906}. 
  The $\chi^2$ fit also provides an estimate $\widehat\sigma_\Omega$
  of the measurement noise. The estimate is used to normalize the
  measured and injected values of $\Omega_0$.}
  \label{fig:measuredvsexpected}
\end{figure}

The same simulated signals can be written to file, and then added
to the interferometer strain channels. These software simulation
signals were added in after the strain data were decimated to
1024~Hz, as shown in Fig.~\ref{fig:dataPipeline}. The flexibility
of software injection allowed a wide range of values for
$\Omega_0\,h_{100}^2$ to be studied. Refer to Table
\ref{tab:signal_injections} for details. This allowed us to follow
the performance of the pipeline to smaller signal-to-noise ratios,
until the signal could no longer be distinguished from the noise.
The behavior of the deduced signal versus injected signal at a
large range of signal-to-noise ratios is presented in
Fig.~\ref{fig:measuredvsexpected}.
%

\section{THE H1-H2 CORRELATION}
\label{sec:H1H2correlation}

The significant instrumental correlation seen between the two LHO
interferometers (H1 and H2) prevents us from establishing an
upper limit on the gravitational wave stochastic background using
what is, potentially, the most sensitive detector pair. It is thus
worth examining this correlation further to understand its
character. We tested the analysis pipeline for contamination from
correlated spectral leakage by re-analyzing the H1-H2 data using
a Hann window on the 90-sec data segments instead of the Tukey
window. The result of this analysis, when scaled for the effective
reduction in observation time, was---within statistical error---the
same as the original Tukey-windowed analysis, discounting this hypothesis.

\begin{figure}[htbp!]
  \includegraphics[width=3.75in,angle=0]{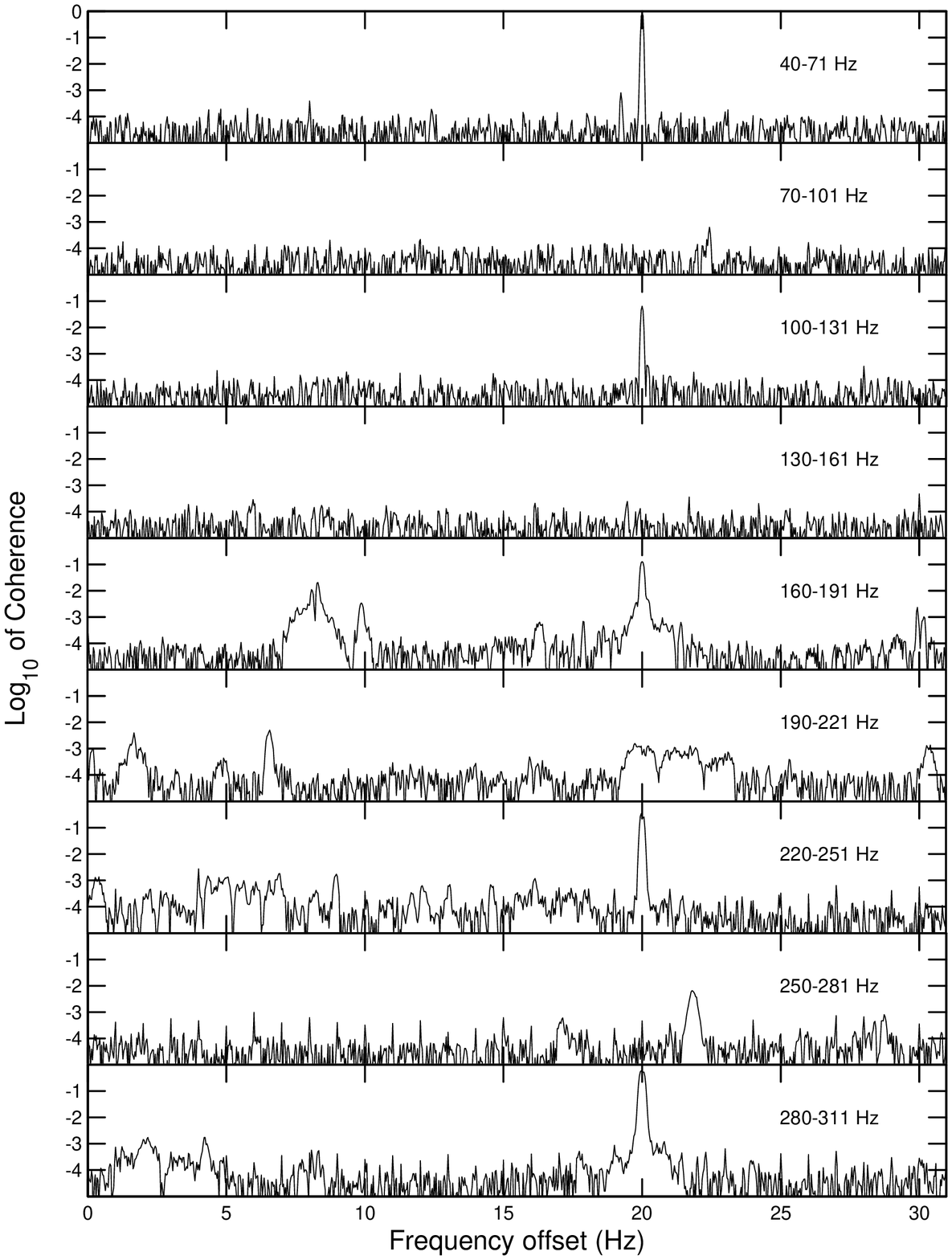}
  \caption{Coherence between the H1 and H2 detector outputs during S1.
  The coherence is calculated with a frequency resolution of 0.033~Hz,
  and with approximately 35,000 periodogram averages; 50\% overlap
  Hann windows were used in the
  Fourier transforms. In addition to the near-unity coherence at 60~Hz
  and harmonics, there is broadened coherence at some of these lines
  due to bilinear mixing of low-frequency seismic noise. There are several
  few-Hz wide regions of significant coherence (around 168~Hz, e.g.), and
  the broad region of significant coherence between 220~Hz and 240~Hz,
  that are likely due to acoustic noise coupling. Also discernible are
  small peaks at many of the integer frequencies between 245-310~Hz,
  likely due to coupling from the GPS 1~pulse-per-second timing signals.}
  \label{fig:h1h2coherence}
\end{figure}

Some likely sources of instrumental correlations are: acoustic
noise coupling to both detectors through the readout hardware
(those components not located inside the vacuum system); common
low-frequency seismic noise that bilinearly mixes with the 60~Hz
and harmonic components, to spill into the analysis band. Figure
\ref{fig:h1h2coherence} shows the coherence function
(Eq.~\ref{e:coherencedef}) between H1 and H2, calculated over
approximately 150 hours of coincident data. It shows signs of
both of these types of sources.

Both of these noise sources are addressable at the instrument
level. Improved electronics equipment being implemented on all
detectors should substantially reduce the $n\cdot 60$~Hz lines,
and consequently the bilinearly mixed sideband components as well.
Better acoustic isolation and control of acoustic sources is also
being planned to reduce this noise source. It is also conceivable
that signal processing techniques, such as those described in
Ref.~\cite{lineremoval, antonyetal}, could be used to remove 
correlated noise, induced by measurable environmental disturbances, 
from the data.

%

\section{CONCLUSIONS AND FUTURE PLANS}

In summary, we have analyzed the first LIGO science data to set
an improved, direct observational upper limit on a stochastic
background of gravitational waves. Our 90\% confidence upper
limit on a stochastic background, having a constant energy
density per logarithmic frequency interval, is
$\Omega_0\,h_{100}^2 \le 23$ in the frequency band $40-314$~Hz.
This is a roughly $10^4$ times improvement over the previous,
broadband interferometric detector measurement.

We described in detail the data analysis pipeline, and tests of
the pipeline using hardware and software injected signals. We
intend to use this pipeline on future LIGO science data to 
set upper limits on $\Omega_0\,h_{100}^2$ at levels which are 
orders of magnitude below unity. 
Two possible additions to the treatment presented here are 
being considered for future analyses: a method for combining 
upper limits from H1-L1 and H2-L1 that takes into account the 
potential H1-H2 instrumental correlations; a Bayesian 
statistical analysis for converting the point estimate into
an upper limit on $\Omega_0\,h_{100}^2$.
Eventually, with both 4-km
interferometers (H1 and L1) operating at the design sensitivity
level shown in Fig.~\ref{fig:sensitivityPlots}, we expect to be
able to set an upper limit using $1$-year of data from this
detector pair at a level $\Omega_0\,h_{100}^2 \le 1 \times 10^{-6}$
in the $40-314$~Hz band. This would improve on the limit from
big-bang nucleosynthesis (see Table~\ref{tab:upperlimits}). The
two interferometers at LHO (H1 and H2) could potentially provide a
lower upper limit, but given our present level of correlated
instrumental noise ($|\widehat\Omega_{\rm inst}\,h_{100}^2| \sim 10$), 
we first need to reduce the correlated noise in each detector by a
factor of $\sim10^4$. We also anticipate cross-correlating L1 with
the ALLEGRO resonant bar detector (located nearby LLO in Baton
Rouge, LA) for a higher frequency search. With this pair
performing at design sensitivity, an upper limit of order
$\Omega_0\,h_{100}^2 \le 0.01$ could be set around 900~Hz, using 1
year of coincident data.

\begin{acknowledgments}
The authors gratefully acknowledge the support of the United States 
National Science Foundation for the construction and operation of 
the LIGO Laboratory and the Particle Physics and Astronomy Research 
Council of the United Kingdom, the Max-Planck-Society and the State 
of Niedersachsen/Germany for support of the construction and operation 
of the GEO600 detector. The authors also gratefully acknowledge the 
support of the research by these agencies and by the Australian 
Research Council, the Natural Sciences and Engineering Research Council 
of Canada, the Council of Scientific and Industrial Research of India, 
the Department of Science and Technology of India, the Spanish 
Ministerio de Ciencia y Tecnologia, the John Simon Guggenheim 
Foundation, the David and Lucile Packard Foundation, the Research 
Corporation, and the Alfred P. Sloan Foundation.
This paper has been assigned LIGO Document Number LIGO-P030009-H-Z.
\end{acknowledgments}

\begin{appendix}
\begin{widetext}
\section{List of symbols}
\label{sec:listofsymbols}


\setlength{\LTcapwidth}{6.25in}
\begin{center}
\begin{longtable}{ p{1.3in} p{4.1in} p{1.3in}}

\caption[Symbols]{A list of symbols that appear in the paper, along 
with their descriptions and equation numbers (if applicable) or 
sections in which they were defined.}
\label{tab:listOfSymbols}
\endlastfoot

Symbol & \multicolumn{1}{c}{Description} &\multicolumn{1}{r} Eq.\ No.\\
\hline

$\Omega_{\rm gw}(f)$ 
& Energy density in gravitational waves per logarithmic frequency 
interval in units of the closure energy density $\rho_{\rm c}$ 
& \multicolumn{1}{r}{\ref{e:Omega_gw}}\\

$\rho_{\rm c}$, $\rho_{\rm gw}$ 
& Critical energy density needed to close the universe, and total 
energy density in gravitational waves 
& \multicolumn{1}{r}{\ref{sec:spectrum}}\\

$h_{100}$, $H_0$, $H_{100}$ 
& $h_{100}$ is the Hubble constant, $H_0$, in units of 
$H_{100}\equiv 100$~km/sec/Mpc 
& \multicolumn{1}{r}{\ref{e:h_100}, \ref{e:H_100}}\\

$h_{ab}(t)$, $h(t)$ 
& Perturbations of the space-time metric, and the corresponding 
gravitational wave strain in a detector 
& \multicolumn{1}{r}{\ref{e:h(t)}}\\

$\vec x_0$, $\widehat{X}^a$, $\widehat{Y}^a$ 
& Position vector of an interferometer vertex, and unit vectors 
pointing in the directions of the arms of an interferometer 
& \multicolumn{1}{r}{\ref{sec:spectrum}}\\

$S_{\rm gw}(f)$ 
& Power spectrum of the gravitational wave strain $h(t)$ 
& \multicolumn{1}{r}{\ref{e:gwpsd}}\\

$s_{i}(t)$, $\widetilde s_i(f)$ 
& Equivalent strain output of the $i$-th detector 
& \multicolumn{1}{r}{\ref{e:s_i(t)}}\\

$h_i(t)$, $\widetilde h_i(f)$ 
& Gravitational wave strain in the $i$-th detector 
& \multicolumn{1}{r}{\ref{e:s_i(t)}}\\

$n_i(t)$, $\widetilde n_i(f)$ 
& Equivalent strain noise in the $i$-th detector 
& \multicolumn{1}{r}{\ref{e:s_i(t)}}\\

$r_i(t)$, $r_i[k]$ 
& Raw (i.e., uncalibrated) output of the $i$-th detector for 
continuous and discrete time
& \multicolumn{1}{r}{\ref{e:r_i[k]}}\\

$\widetilde R_i(f)$, $\widetilde R_{iI}[\ell]$ 
& Response function for the $i$-th detector, and the discrete
frequency response function for interval $I$
& \multicolumn{1}{r}{\ref{sec:pipeline}}\\

$t_k, f_{\ell}$
& Discrete time and discrete frequency values
& \multicolumn{1}{r}{\ref{sec:pipeline}}\\

$\delta t, \delta f$
& Sampling period of the time-series data 
($1/1024$~sec after down-sampling), and bin spacing (0.25~Hz) of 
the discrete power spectra, optimal filter, $\ldots$
& \multicolumn{1}{r}{\ref{sec:pipeline}}\\

$\Delta f$
& General frequency resolution of discrete Fourier transformed data
& \multicolumn{1}{r}{\ref{sec:freqselection}}\\

$N$ 
& Number of discrete-time data points in one segment of data 
& \multicolumn{1}{r}{\ref{sec:pipeline}}\\

$r_{iIJ}[k]$,
$g_{iIJ}[k]$,
$\widetilde g_{iJK}[q]$ 
& Raw detector output for the $J$-th segment in interval $I$ 
evaluated at discrete time $t_k$, and the corresponding windowed 
and zero-padded time series and discrete Fourier transform 
& \multicolumn{1}{r}{\ref{e:g_iIJ}}\\

$\mathcal{G}_{IJ}[\ell]$
& Cross-spectrum of the windowed and zero-padded raw time-series,
binned to match the frequency resolution of the optimal filter 
$\widetilde Q_I[\ell]$
& \multicolumn{1}{r}{\ref{e:G_12IJ}}\\

$w_i[k]$ 
& Window function for the $i$-th detector 
& \multicolumn{1}{r}{\ref{sec:pipeline}}\\

$n_b$ 
& Number of frequency values binned together to match the 
frequency resolution of the optimal filter $\widetilde Q_I[\ell]$
& \multicolumn{1}{r}{\ref{sec:pipeline}}\\

$\ell_{\rm min}$, $\ell_{\rm max}$
& Indices corresponding to the maximum and minimum frequencies 
used in the calculation of the cross-correlation $Y_{IJ}$
& \multicolumn{1}{r}{\ref{sec:pipeline}}\\

$\delta_T(f)$ 
& Finite-time approximation to the Dirac delta function $\delta(f)$ 
& \multicolumn{1}{r}{\ref{e:delta_T}}\\

$T, T_{\rm seg}$, $T_{\rm int}$ 
& General observation time, and durations of an individual data segment 
and interval (90 sec, 900 sec)
& \multicolumn{1}{r}{\ref{sec:pipeline}}\\

$Y$ 
& General cross correlation of two detectors
& \multicolumn{1}{r}{\ref{e:Y_continuous_time}}\\

$Q(t)$, $\widetilde{Q}(f)$ 
& Optimal filter for the cross-correlation $Y$ 
& \multicolumn{1}{r}{\ref{e:optimal_continuous}}\\

$\mu_Y$, $\sigma^2_Y$, $\rho_Y$ 
& Theoretical mean, variance, and signal-to-noise ratio of the 
cross-correlation $Y$ 
& \multicolumn{1}{r}{\ref{e:mu_continuous},\ref{e:sigma2_continuous}}\\

$\langle\rho_Y\rangle$ 
& Expected value of the signal-to-noise ratio of the 
cross-correlation $Y$ 
& \multicolumn{1}{r}{\ref{e:SNR}}\\

$Y_{IJ}$, $Y_I$
& Cross-correlation for the $J$-th segment in interval $I$,
and average of the $Y_{IJ}$ 
& \multicolumn{1}{r}{\ref{e:Y_discrete_freq}, \ref{e:Y_I}}\\

$Y_{\rm opt}$ 
& Weighted average of the $Y_I$ 
& \multicolumn{1}{r}{\ref{e:Yoptimal}}\\

$x_{IJ}$
& Cross-correlation values $Y_{IJ}$ with mean removed and normalized
by the theoretical variances
& \multicolumn{1}{r}{\ref{sec:results}}\\

$\widetilde Y_{IJ}[\ell]$, $\widetilde Y_{I}[\ell]$, 
$\widetilde Y_{\rm opt}[\ell]$ 
& Summands of $Y_{IJ}$, $Y_I$, $Y_{\rm opt}$ 
& \multicolumn{1}{r}{\ref{e:Y_IJ[l]}, \ref{e:Y_I[l]}, \ref{e:Y_opt[l]}}\\

$\widetilde Q_I[\ell]$ 
& Optimal filter for the cross-correlation $Y_{IJ}$ 
& \multicolumn{1}{r}{\ref{e:optimal_discrete}}\\

$\gamma(f)$, $\gamma[\ell]$ 
& Overlap reduction function evaluated at frequency $f$, and 
discrete frequency $f_{\ell}$ 
& \multicolumn{1}{r}{\ref{sec:ccstatistic}, \ref{sec:pipeline}}\\

$\gamma_{\rm rms}$, $\Delta_{\rm BW}$
& Root-mean-square value of the overlap reduction function over the
corresponding frequency bandwidth $\Delta_{\rm BW}$
& \multicolumn{1}{r}{\ref{sec:ccstatistic}}\\

$P_i(f)$, $P_{iI}[l]$ 
& Power spectrum of the strain noise in the $i$-th detector, and 
the discrete frequency strain noise power spectrum estimate 
for interval $I$ 
& \multicolumn{1}{r}{\ref{sec:ccstatistic}, \ref{sec:pipeline}}\\

${\cal N}$, ${\cal N}_I$ 
& Normalization factors for the optimal filter $\widetilde Q$, 
$\widetilde Q_I$ 
& \multicolumn{1}{r}{\ref{e:normalization_continuous}, 
\ref{e:normalization_discrete}}\\

$\overline{w_1w_2}$, $\overline{w_1^2w_2^2}$ 
& Overall multiplicative factors introduced by windowing 
& \multicolumn{1}{r}{\ref{e:window_fac1}, \ref{e:window_fac2}}\\

$\sigma^2_{Y_{IJ}}$, $s^2_{Y_{IJ}}$ 
& Theoretical and estimated variance of the cross-correlation $Y_{IJ}$ 
& \multicolumn{1}{r}{\ref{e:sigma2_discrete_final}, \ref{e:s_I}}\\

$\sigma^2_{Y_I}$, $\sigma^2_{Y_{\rm opt}}$ 
& Theoretical variance of $Y_I$ and $Y_{\rm opt}$ 
& \multicolumn{1}{r}{\ref{sec:errors}}\\

$\mu_{\widetilde Y_{\rm opt}}[\ell]$, 
$\sigma^2_{\widetilde Y_{\rm opt}}[\ell]$ 
& Theoretical mean and variance of $\widetilde Y_{\rm opt}[\ell]$ 
& \multicolumn{1}{r}{\ref{e:mu_Yopt}, \ref{e:sigma2_Yopt}}\\

$P_{12}(f)$
& Cross-power spectrum of the strain noise between two detectors
& \multicolumn{1}{r}{\ref{sec:freqselection}}\\
 
$\Gamma_{12}(f)$ 
& Coherence function between two detectors 
& \multicolumn{1}{r}{\ref{e:coherencedef}}\\

$N_{\rm avg}$, $\sigma_\Gamma$
& Number of averages used in the measurement of the coherence, and
the corresponding statistical uncertainity in the measurement
& \multicolumn{1}{r}{\ref{sec:freqselection}}\\

$\Omega_0$, ${\widehat{\Omega}}_0$ 
& Actual and estimated values of an (assumed) constant value of 
$\Omega_{\rm gw}(f)$ due to gravitational waves 
& \multicolumn{1}{r}{\ref{sec:spectrum}, \ref{sec:pipeline}}\\

$\Omega_{\rm inst}$, ${\widehat{\Omega}}_{\rm inst}$ 
& Actual and estimated values of the instrumental contribution to 
the measured cross-correlation 
& \multicolumn{1}{r}{\ref{sec:results}}\\

$\Omega_{\rm eff}$, ${\widehat{\Omega}}_{\rm eff}$ 
& Actual and estimated values of an effective $\Omega$ due to 
instrumental and gravitational wave effects 
& \multicolumn{1}{r}{\ref{e:hat_Omega_eff}}\\

$\sigma^2_{\Omega}$, ${\widehat{\sigma}}^2_{\Omega}$ 
& Actual and estimated variances of $\widehat\Omega_{\rm eff}$ due 
to statistical variations in $Y_{\rm opt}$ 
& \multicolumn{1}{r}{\ref{e:sigma_Omega}, \ref{e:hat_sigma_Omega}}\\

${\sigma}^2_{\Omega,{\rm cal}}$, 
${\widehat{\sigma}}^2_{\Omega,{\rm cal}}$ 
& Actual and estimated variances of $\widehat\Omega_{\rm eff}$ due 
to variations in the instrument calibration 
& \multicolumn{1}{r}{\ref{sec:errors}}\\

${\sigma}^2_{\Omega,{\rm psd}}$, 
${\widehat{\sigma}}^2_{\Omega,{\rm psd}}$ 
& Actual and estimated variances of $\widehat\Omega_{\rm eff}$ due 
to variations in the noise power spectra 
& \multicolumn{1}{r}{\ref{sec:errors}}\\

${\widehat{\sigma}}^2_{\Omega,{\rm tot}}$ 
& Estimated variance of $\widehat\Omega_{\rm eff}$ due 
to combined statistical, calibration, and power spectra variations 
& \multicolumn{1}{r}{\ref{e:hat_sigma2_Omega_tot}}\\

$\chi^2(\Omega_0)$
& Chi-squared statistic to compare $\widetilde Y_{\rm opt}[\ell]$ to its 
expected value for a stochastic background with 
$\Omega_{\rm gw}(f)\equiv \Omega_0$
& \multicolumn{1}{r}{\ref{e:chisquare}}\\ 

$\chi^2_{\rm min}$
& Minimum chi-squared value per degree of freedon
& \multicolumn{1}{r}{\ref{sec:fdomain}}\\

$Y(\tau)$, $\widetilde Y(f)$
& Cross correlation statistic as a function of time-shift $\tau$, and
its Fourier transform
& \multicolumn{1}{r}{\ref{e:YofTau}, \ref{e:Yoff}}\\
\\
\\
\end{longtable}
\end{center}


\end{widetext}
\end{appendix}


\end{document}